\documentclass[11pt,a4paper]{article} 
\pdfoutput=1

\usepackage{jcapmod} 
\usepackage{booktabs}
\usepackage{url}
\usepackage[utf8]{inputenc}
\usepackage{enumitem}
\usepackage{amsmath}

\newcommand{\eq}[1]{eq. (\ref{#1})} 
\newcommand{\fig}[1]{figure \ref{#1}}
\newcommand{\sektion}[1]{section \ref{#1}}

\newcommand{\Sektion}[1]{Section \ref{#1}}

\def\ud{\mathrm{d}}

\definecolor{forfootnote}{RGB}{200,0,255}
\makeatletter
\renewcommand\@makefnmark{\hbox{\@textsuperscript{\normalfont\color{forfootnote}\@thefnmark}}}
\renewcommand\@makefntext[1]{%
  \parindent 1em\noindent
            \hb@xt@1.8em{%
                \hss\@textsuperscript{\normalfont\@thefnmark}}#1}
\makeatother

\title{Inflationary predictions of double-well, Coleman-Weinberg, and
hilltop potentials with non-minimal coupling}

\author{Nilay Bostan,}
\author{\"Omer G\"ulery\"uz}
\author[*]{and Vedat Nefer \c{S}eno\u{g}uz%
\note[*]{Corresponding author.}}

\affiliation{Department of Physics, Mimar Sinan Fine Arts University, \\ 34380
\c{S}i\c{s}li, \.Istanbul, Turkey}

\emailAdd{nilay.bostan@yeditepe.edu.tr}
\emailAdd{guleryuz17@itu.edu.tr}
\emailAdd{nefer.senoguz@msgsu.edu.tr}

\abstract{We discuss how the non-minimal coupling $\xi\phi^2R$ between the
inflaton and the Ricci scalar affects the predictions of single field
inflation models where the inflaton has a non-zero vacuum expectation value
(VEV) $v$ after inflation. We show that, for inflaton values both above the
VEV and below the VEV during inflation, under certain conditions the
inflationary predictions become approximately the same as the predictions
of the Starobinsky model.  We then analyze inflation with double-well and
Coleman-Weinberg potentials in detail, displaying the regions in the
$v$-$\xi$ plane for which the spectral index $n_s$ and the tensor-to-scalar
ratio $r$ values are compatible with the current observations. $r$ is
always larger than $0.002$ in these regions.  Finally, we consider the
effect of $\xi$ on small field inflation (hilltop) potentials.}


\arxivnumber{1802.04160}

\begin{document} 

%comment out the line below before submitting to arXiv
%\begin{flushright}\today\end{flushright}

\maketitle \flushbottom

%adjust distance before and after equation
\setlength{\belowdisplayskip}{7.5pt} \setlength{\belowdisplayshortskip}{7.5pt}
\setlength{\abovedisplayskip}{7.5pt} \setlength{\abovedisplayshortskip}{7.5pt}

%%%%%%%%%%%%%%%%%%%%%%%%%%%%%%%%%%%%%%%%%%%%%%%%%%%%%%%%%%%%%%%%

\section{Introduction} \label{intro}

The hypothesis of cosmic inflation
\cite{Guth:1980zm,Linde:1981mu,Albrecht:1982wi,Linde:1983gd} provides a
plausible explanation of the large scale homogeneity of the universe and,
more importantly, of the primordial density perturbations that evolve into
cosmic structure. A simple way inflation can occur is based on a
slow-rolling scalar field $\phi$ called the inflaton. Once the Lagrangian
for the inflaton field and also the thermal history of the universe after
inflation is specified, values for observational parameters can be
calculated and compared with constraints coming from measurements of the
cosmic microwave background (CMB) anisotropies
\cite{Ade:2015xua,Ade:2015lrj}.

The observational parameters, in particular the scalar spectral index $n_s$
and the tensor-to-scalar ratio $r$, have been calculated for various
inflationary potentials (see \cite{Martin:2013tda} for a comprehensive
subset).  An assumption often made in the calculations is that the inflaton
is minimally coupled. On the other hand, a renormalizable scalar field
theory in curved space-time also requires the non-minimal coupling
$\xi\phi^2R$ between the inflaton and the Ricci scalar
\cite{Callan:1970ze,Freedman:1974ze,Buchbinder:1992rb}.
For a given potential, depending on the value of the non-minimal coupling
parameter $\xi$, inflationary predictions and even whether inflation occurs
or not can change
\cite{Abbott:1981rg,Spokoiny:1984bd,Lucchin:1985ip,Futamase:1987ua,Fakir:1990eg,Salopek:1988qh,Amendola:1990nn,Faraoni:1996rf,Faraoni:2004pi}.

Here we will investigate how the value of the non-minimal coupling
parameter $\xi$ affects the inflationary predictions for potentials where
the inflaton has a non-zero vacuum expectation value (VEV) $v$ after
inflation. In terms of the redefined field $\varphi\equiv\phi-v$, the
non-minimal coupling in the Lagrangian includes a linear term in $\varphi$
as well as a quadratic term. Under some conditions on $\xi$ and $v$ that
are discussed in \sektion{nonminimal}, this leads to inflationary
predictions that approach those of the Starobinsky ($R^2$ inflation) model
\cite{Starobinsky:1980te}, which is in good agreement with the current
observations \cite{Ade:2015lrj}.  The Starobinsky-like behaviour is
obtained not just for the well-known non-minimally coupled quartic
potential case but also when inflation occurs near the quadratic minimum of
the potential, for inflaton values above (below) the VEV and $\xi>0$
($\xi<0$).

A reason for considering a non-zero VEV after inflation is that such
potentials can be associated with symmetry breaking in the early universe.
After a general discussion of inflation with non-minimal coupling for such
potentials, we then analyze in detail two archetypal symmetry breaking
potentials, namely the double-well potential (\sektion{double}) and the
Coleman-Weinberg potential (\sektion{cw}).  Although both potentials with
non-minimal coupling were previously considered, there are some gaps and
disagreements in the literature which we address in these sections. For
each potential, we display the observational parameter values as functions of
$v$ for selected $\xi$ values as well as the regions in the $v$-$\xi$
plane for which the spectral index $n_s$ and tensor-to-scalar ratio $r$
values are compatible with the current observations. \Sektion{small}
suggests modifying the double-well potential to obtain a small field
inflation (hilltop) potential, which unlike the other two potentials can
fit observations for inflaton values below the VEV and $\xi,\,v\ll1$. Finally,
\sektion{conclude} concludes the paper with a summary of our results and a
remark on perturbative unitarity violation.

It is worth mentioning that we use the metric formulation of gravity
throughout the paper. For inflation with a non-minimally coupled scalar
field, the Palatini formulation leads to different predictions for
cosmological parameters \cite{Bauer:2008zj}. In particular, the attractor
behaviour leading to the predictions of the Starobinsky model is lost, and
$r$ can be much smaller compared to the metric formulation
\cite{Bauer:2008zj,Jarv:2017azx}.

%%%%%%%%%%%%%%%%%%%%%%%%%%%%%%%%%%%%%%%%%%%%%%%%%%%%%%%%%%%%%%%%

\section{Inflation with non-minimal coupling} \label{nonminimal}

Suppose we have a non-minimally coupled scalar field $\phi$ with a canonical kinetic
term and a potential $V_J(\phi)$:
\begin{equation} \label{vjphi}
\frac{\mathcal{L}_J}{\sqrt{-g}}=\frac12F(\phi)R-\frac12g^{\mu\nu}\partial_{\mu}\phi\partial_{\nu}\phi-V_J(\phi)\,,
\end{equation}
where the subscript $J$ indicates that the Lagrangian is specified in a
Jordan frame. Here, for $V_J(\phi)$ we will be considering
symmetry-breaking type of potentials where the inflaton $\phi$ takes
positive values and has a non-zero vacuum expectation value (VEV) $v$ after
inflation. 

Our choice for $F(\phi)$ consists of a constant $m^2$ term and a
non-minimal coupling $\xi\phi^2R$ between the inflaton and the Ricci
scalar. The constant term is familiar from the Einstein-Hilbert action,
and the $\xi\phi^2R$ term is required in a renormalizable scalar field
theory in curved space-time
\cite{Callan:1970ze,Freedman:1974ze,Buchbinder:1992rb}. We
are using units where the reduced Planck scale $m_P=1/\sqrt{8\pi
G}\approx2.4\times10^{18}\text{ GeV}$ is set equal to unity, so we require
$F(\phi)\to1$ after inflation. Therefore taking $m^2=1-\xi v^2$,
we have $F(\phi)=m^2+\xi\phi^2=1+\xi(\phi^2-v^2)$.
The $\xi v^2$ term in $F(\phi)$ can be neglected in some specific models
such as when the standard model Higgs is the inflaton
\cite{Bezrukov:2007ep,Atkins:2012yn}, but may well play an important role
in other models. For instance, if inflation is associated with symmetry
breaking at or near the grand unified theory scale $v\sim0.01$,
$|\xi|v^2\gtrsim1$ is possible for values of $|\xi|\gtrsim10^4$ similar to
values required for standard model Higgs inflation.

As we will see,  as cosmological scales exit the horizon $F(\phi)\gtrsim1$
in the observationally favored region of parameters, and the
Starobinsky-like regime corresponds to $F(\phi)\gg1$. The effective
gravitational constant $G_N=1/[8\pi F(\phi)]$ remains positive throughout
the evolution of the field. Indeed, if we switch to the Einstein frame and
make a field redefinition so that the kinetic term is again canonical, we
see that $G_N=0$ is only reached at infinite values of the field (see
\sektion{calculate} and ref. \cite{Linde:2011nh}). This implies, in
particular, that if $\xi v^2>1$ there can be no transition from the
symmetric ($\phi=0$) phase to the broken-symmetry ($\phi=v$) phase.
Nevertheless, we include this case in our investigations as the field
evolution does not have to start from the symmetric phase, and could for
example start from values above the VEV as would be expected for chaotic
initial conditions \cite{Linde:1983gd}.

It has been appreciated
\cite{Whitt:1984pd,Salopek:1988qh,Barbon:2009ya,Linde:2011nh,Kallosh:2013hoa,Kallosh:2013maa,Kallosh:2013tua,Kehagias:2013mya,Giudice:2014toa,Galante:2014ifa}
that different choices for $F(\phi)$ and $V_J(\phi)$ can share the same attractor
point with the Starobinsky ($R^2$ inflation) model
\cite{Starobinsky:1980te,Mukhanov:1981xt}, which predicts
\begin{equation}\label{starpoint}
n_s=1-\frac{2}{N}\,,\quad r=\frac{12}{N^2}\,,\quad \frac{\ud n_s}{\ud\ln k}=-\frac{2}{N^2}\,,
\end{equation}
to leading order in the number of e-folds $N$, where 
$\mathrm{d} n_s/\mathrm{d} \ln k$ is the running of the
spectral index. An example relevant to our discussion is
\begin{equation}\label{3950}
F(\phi)=1+\xi \phi^n\,,\qquad V_J(\phi)\propto \phi^{2n}\,,
\end{equation}
which is a special case of the strong coupling attractor model
discussed in \cite{Kallosh:2013tua} (see also \cite{Barbon:2009ya}).

In terms of the redefined field $\varphi\equiv\phi-v$ so that $\varphi=0$
after inflation, $F(\varphi)=1+\xi\varphi^2(1+2v/\varphi)$ includes a
linear term in $\varphi$ as well as a quadratic term. If $\phi^2-v^2\gg
v^2$ ($\varphi^2\gg v^2$) as cosmological scales exit the horizon, it means
the inflaton is away from the minimum and
$F(\varphi)\approx1+\xi\varphi^2$. Then \eq{3950} is satisfied for
$V_J(\varphi)\propto\varphi^4$, the non-minimally coupled quartic model
well-known since the late eighties
\cite{Fakir:1990eg,Salopek:1988qh,Okada:2010jf,Bezrukov:2013fca}. On the
other hand, if $|\phi^2-v^2|\ll v^2$ ($\varphi^2\ll v^2$) as cosmological
scales exit the horizon, $F(\varphi)\approx1+2\xi v\varphi$ so that
\eq{3950} is satisfied for $V_J(\varphi)\propto\varphi^2$, with $\xi>0$
($\xi<0$) for inflaton values above (below) the VEV during inflation. 

Since a generic potential will be quadratic close enough to its minimum, it
seems that a generic $V_J(\phi)$ can share the predictions of the
Starobinsky model up to leading order in the number of e-folds $N$. For
this to happen, $\xi$ and $v$ values should satisfy some constraints which
we discuss in \sektion{star}.  Before that, we briefly review how to
calculate the observational parameters for inflation with non-minimal coupling.

%%%%%%%%%%%%%%%%%%%%%%%%%%%%%%%%%%%%%%%%%%%%%%%%%%%%%%%%%%%%%%%%

\subsection{Calculating the observational parameters} \label{calculate}
For calculating the observational parameters given \eq{vjphi}, it is convenient
to switch to the Einstein ($E$) frame by applying a Weyl rescaling
$g_{\mu\nu}=\tilde{g}_{\mu\nu}/F(\phi)$, so that the Lagrangian density
takes the form \cite{Fujii:2003pa}
\begin{equation} \label{LE}
\frac{\mathcal{L}_E}{\sqrt{-\tilde{g}}}=\frac12\tilde{R}-\frac{1}{2Z(\phi)}\tilde{g}^{\mu\nu}\partial_{\mu}\phi\partial_{\nu}\phi-V_E(\phi)\,,
\end{equation}
where
\begin{equation} \label{Zphi}
\frac{1}{Z(\phi)}=\frac32\frac{F'(\phi)^2}{F(\phi)^2}+\frac{1}{F(\phi)}\,,\qquad
V_E(\phi)=\frac{V_J(\phi)}{F(\phi)^2}\,,
\end{equation}
and $F'\equiv\ud F/\ud\phi$. If we make a field redefinition
\begin{equation}\label{redefine}
\ud\sigma=\frac{\ud\phi}{\sqrt{Z(\phi)}}\,,
\end{equation}
we obtain the Lagrangian density for a
minimally coupled scalar field $\sigma$ with a canonical kinetic term. 

For $F(\phi)=1+\xi(\phi^2-v^2)$, \eq{Zphi} gives
\begin{equation} \label{Zphiexplicit}
\frac{1}{Z(\phi)}=\frac{1+\xi(\phi^2-v^2)+6\xi^2\phi^2}{\left[1+\xi(\phi^2-v^2)\right]^2}\,.
\end{equation}
It will be useful to consider some simplifying cases of this expression: 
\begin{description}[style=multiline,font=\normalfont]
\item[1.] Weak coupling limit \\ 
If $|\xi(\phi^2-v^2)|\ll1$ and $6\xi^2\phi^2\ll1$, $\phi\approx\sigma$ and
$V_J(\phi)\approx V_E(\sigma)$. (Provided $|\xi|\ll1/6$, these conditions
will be satisfied when $|\xi|v^2\ll1$ for inflation below the VEV, and
$|\xi|\phi^2\ll1$ for inflation above the VEV.) Then, the inflationary
predictions are approximately the same as for minimal coupling in general.
Note, however, that if $V_J(\phi)$ is very flat as cosmological scales exit
the horizon, then even a small correction in the potential can
significantly alter the inflationary predictions, as we will discuss in
\sektion{small}.
\item[2.] Induced gravity limit \cite{Zee:1978wi} \\
In this limit ($\xi v^2=1$, $F(\phi)=\xi\phi^2$), \eq{Zphiexplicit} simplifies to
$Z(\phi)=\xi\phi^2/(1+6\xi)$ and using \eq{redefine}, we obtain
\begin{equation}\label{induced}
\phi=v\exp\left(\sqrt{\frac{\xi}{1+6\xi}}\sigma\right)\,,
\end{equation}
where we took $\sigma(v)=0$.
\item[3.] Strong coupling limit \\
If $6\xi^2\phi^2\gg|\xi(\phi^2-v^2)|\gg1$, we have
\begin{equation} \label{strong}
\frac{1}{Z(\phi)}\approx\frac{6\phi^2}{(\phi^2-v^2)^2}\,.
\end{equation}
Using \eq{redefine}, we obtain
$\phi^2-v^2\propto e^{2\sigma/\sqrt6}$ where $\sigma$ is
positive during inflation. This exponential behaviour in
terms of the canonical field $\sigma$ makes it difficult to satisfy
observations except for the special cases discussed in \sektion{star} where the
Einstein frame potential $V_E(\sigma)$ has a plateau due to cancellations
between $V_J(\phi)$ and $F(\phi)^2$.  \end{description}

Once the Einstein frame potential is expressed in terms of the canonical
$\sigma$ field, the observational parameters can be calculated using the
slow-roll parameters (see ref. \cite{Lyth:2009zz} for a review and
references):
\begin{equation}\label{slowroll1}
\epsilon =\frac{1}{2}\left( \frac{V_{\sigma} }{V}\right) ^{2}\,, \quad
\eta = \frac{V_{\sigma \sigma} }{V}  \,, \quad
\xi ^{2} = \frac{V_{\sigma} V_{\sigma \sigma\sigma} }{V^{2}}\,,
\end{equation}
where $\sigma$'s in the subscript denote derivatives.
The spectral index
$n_s$, the tensor-to-scalar ratio
$r$ and the running of the spectral index
$\mathrm{d} n_s/\mathrm{d} \ln k$ are given in the slow-roll
approximation by
\begin{equation}\label{nsralpha1}
n_s = 1 - 6 \epsilon + 2 \eta \,,\quad
r = 16 \epsilon \,,\quad
\frac{\ud n_s}{\ud\ln k} = 16 \epsilon \eta - 24 \epsilon^2 - 2 \xi^2\,.
\end{equation}

The amplitude of the curvature perturbation $\Delta_\mathcal{R}$ is given by
\begin{equation} \label{perturb1}
\Delta_\mathcal{R}=\frac{1}{2\sqrt{3}\pi}\frac{V^{3/2}}{|V_{\sigma}|}\,,
\end{equation}
which should satisfy $\Delta_\mathcal{R}^2\approx 2.4\times10^{-9}$ from
the Planck measurement \cite{Ade:2015xua} with the pivot scale chosen at
$k_* = 0.002$ Mpc$^{-1}$. The number of e-folds is given by
\begin{equation} \label{efold1}
N_*=\int^{\sigma_*}_{\sigma_e}\frac{V\rm{d}\sigma}{V_{\sigma}}\,, \end{equation}
where the subscript ``$_*$'' denotes quantities when the scale
corresponding to $k_*$ exited the horizon, and $\sigma_e$ is the inflaton
value at the end of inflation, which we estimate by $\epsilon(\sigma_e) =
1$.

Unfortunately, for general values of $\xi$ and $v$, it is difficult and
inconvenient to express the potential in terms of the canonical field
$\sigma$.  We therefore rewrite these slow-roll expressions in terms of the
original field $\phi$ for the numerical calculations, following the
approach in ref. \cite{Linde:2011nh}. Using \eq{redefine},
\eq{slowroll1} can be written as
\begin{equation}\label{slowroll2}  
\epsilon=Z\epsilon_{\phi}\,,\quad
\eta=Z\eta_{\phi}+{\rm sgn}(V')Z'\sqrt{\frac{\epsilon_{\phi}}{2}}\,,\quad
\xi^2=Z\left(Z\xi^2_{\phi}+3{\rm sgn}(V')Z'\eta_{\phi}\sqrt{\frac{\epsilon_{\phi}}{2}}+Z''\epsilon_{\phi}\right)\,.
\end{equation}
where we defined 
\begin{equation}
\epsilon_{\phi} =\frac{1}{2}\left( \frac{V^{\prime} }{V}\right) ^{2}\,, \quad
\eta_{\phi} = \frac{V^{\prime \prime} }{V}  \,, \quad
\xi ^{2} _{\phi}= \frac{V^{\prime} V^{\prime \prime\prime} }{V^{2}}\,.
\end{equation}
Similarly, \eq{perturb1} and \eq{efold1} can be written as
\begin{eqnarray}\label{perturb2}
\Delta_\mathcal{R}&=&\frac{1}{2\sqrt{3}\pi}\frac{V^{3/2}}{\sqrt{Z}|V^{\prime}|}\,,\\
\label{efold2} N_*&=&\rm{sgn}(V')\int^{\phi_*}_{\phi_e}\frac{\ud\phi}{Z(\phi)\sqrt{2\epsilon_{\phi}}}\,.
\end{eqnarray}

To calculate the numerical values of $n_s$, $r$ and $\alpha$ we also need a
numerical value of $N_*$. Assuming a standard thermal history after inflation, 
\begin{equation} \label{efolds}
N_*\approx64.7+\frac12\ln\frac{\rho_*}{m^4_P}-\frac{1}{3(1+\omega_r)}\ln\frac{\rho_e}{m^4_P}
+\left(\frac{1}{3(1+\omega_r)}-\frac14\right)\ln\frac{\rho_r}{m^4_P}\,.
\end{equation}
Here $\rho_{e}=(3/2)V(\phi_{e})$ is the
energy density at the end of inflation, $\rho_r$ is the energy density at the
end of reheating and $\omega_r$ is the equation of state parameter during
reheating, which we take to be constant.\footnote{For a derivation of
\eq{efolds} see e.g. ref. \cite{Liddle:2003as}. Note that $N_*$ is defined
in the Einstein frame. The number of e-folds in the Jordan frame
$N_*^J=N_*+(1/2)\ln[F(\phi_*)/F(\phi_e)]$ can be noticeably different in
the strong coupling limit \cite{Lerner:2009xg,Lerner:2009na,Burns:2016ric,Karam:2017zno}. However, in a Jordan frame calculation
the additional term in $N_*^J$ would appear in both eqs. \ref{efold2} and
\ref{efolds}, leaving the physically observable quantities unchanged, as
expected \cite{Postma:2014vaa}.} Using \eq{perturb1}, we can express $\rho_*$ in terms of $r$:
\begin{equation}\label{nstarandr}
\rho_*\approx V(\phi_*)=\frac{3\pi^2\Delta_\mathcal{R}^2 r}{2}\,.
\end{equation}

\begin{figure}[!t]
\centering
\includegraphics[angle=0, width=10cm]{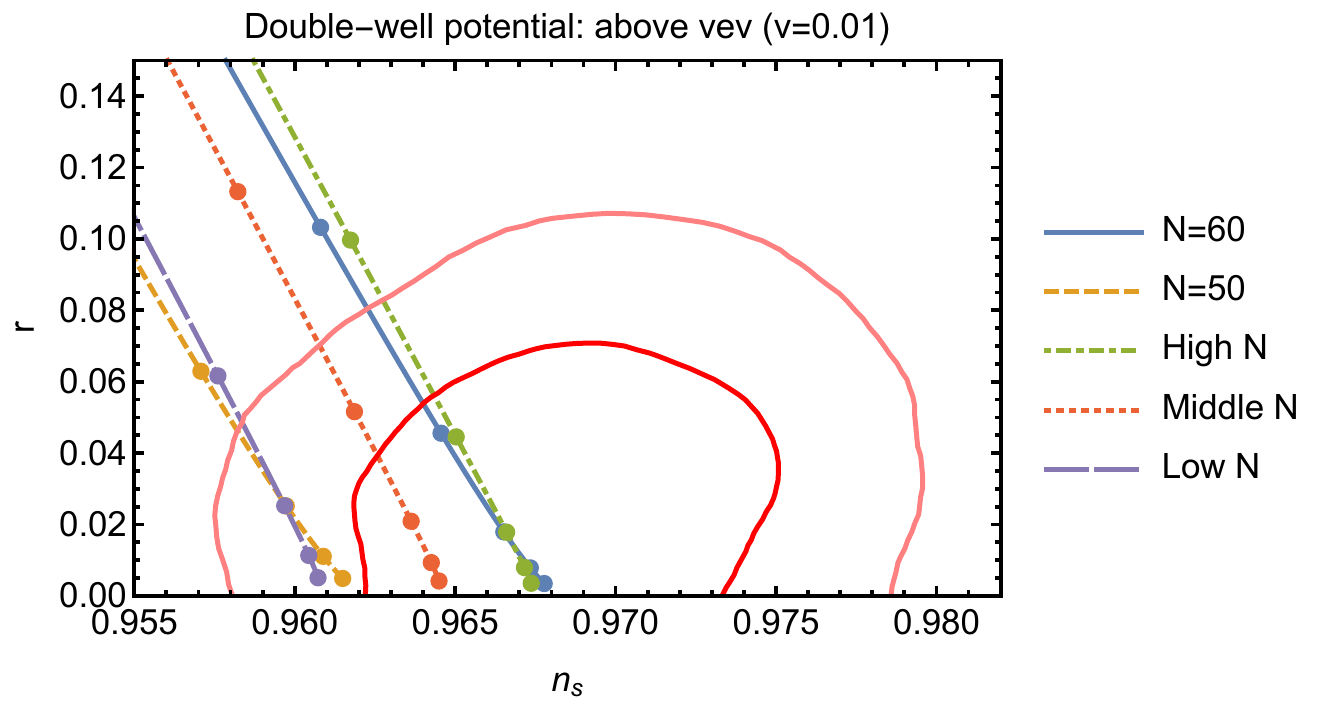}
\caption{$n_s$-$r$ predictions for varying $\xi$ values and different reheating
scenarios as explained in the text. The dots on the curves correspond to
$\xi=10^{-2.5},\,10^{-2},\,10^{-1.5},\,0.1,$ and $\,1$, top to bottom. The pink
(red) contour corresponds to the 95\% (68\%) CL contour
 given by the Planck collaboration (Planck TT+lowP+BKP+lensing+ext) \cite{Ade:2015xua}.}
  \label{higgs_above_reheat}
\end{figure}

To represent a plausible range of $N_*$, we can consider three cases: In
the high-$N$ case $\omega_{r}$ is taken to be 1/3, which is equivalent to
assuming instant reheating.  In the middle-$N$ case we take $\omega_{r}=0$
and the reheat temperature $T_r=10^9$ GeV, calculating $\rho_r$ using the
standard model value for the number of relativistic degrees of freedom ($g_*=106.75$).
In the low-$N$ case we take $T_r=100$ GeV (again with
$\omega_{r}=0$).\footnote{$T_r$ as low as 10 MeV is consistent with big
bang nucleosynthesis, however it is difficult to explain how baryogenesis
could occur at such low temperatures.} The $n_s$ vs.  $r$ curve for each
case is shown in figure \ref{higgs_above_reheat} for the double-well
potential (discussed in \sektion{double}) along  with the 
68\% and 95\% confidence level (CL) contours given by the Planck collaboration
(Planck TT+lowP+BKP+lensing+ext) \cite{Ade:2015xua}. The figure shows that
for the double-well potential, the fiducial $N_*$ values of 50 and 60 that
are often used essentially coincide with the range expected from a standard
thermal history after inflation. This is also the case for the
Coleman-Weinberg potential discussed in \sektion{cw}. However, $N_*$ is
smaller (e.g. between approximately 45 and 55 if $v\sim0.01$) for the small
field inflation models discussed in \sektion{small} due to inflation
occurring at a lower energy scale.

We have carried out all the calculations in this article up to the leading
order in the slow-roll parameters. Higher order corrections slightly change
the values of the observational parameters (see e.g. refs.
\cite{Lyth:2009zz,Karam:2017zno}). However, the uncertainty in the values
of these parameters due to the reheating stage is much larger compared to
the theoretical errors associated with the slow-roll approximation. 

%%%%%%%%%%%%%%%%%%%%%%%%%%%%%%%%%%%%%%%%%%%%%%%%%%%%%%%%%%%%%%%%

\subsection{The Starobinsky conditions} \label{star}

As mentioned in the beginning of \sektion{nonminimal}, for a potential
$V_J(\varphi)$ which is quartic away from the minimum or quadratic close to
the minimum, if some conditions on $\xi$ and $v$ values are satisfied,
predictions approach the Starobinsky point given by \eq{starpoint} on the
$n_s$-$r$  plane.  Following the discussion in ref.
\cite{Galante:2014ifa}, we will now derive these conditions using the
relation of the Starobinsky point with the order and residue of the leading
pole in the kinetic term.

Let's write the Einstein frame Lagrangian density in terms of $\chi(\phi)\equiv 1/F(\phi)$:
\begin{equation} \label{LEchi}
\frac{\mathcal{L}_E}{\sqrt{-\tilde{g}}}=\frac12\tilde{R}-\frac{1}{2}K(\chi)\tilde{g}^{\mu\nu}\partial_{\mu}\chi\partial_{\nu}\chi-V_E(\chi)\,.
\end{equation}
Suppose $K(\chi)$ is given by a Laurent series with a leading pole located at
$\chi=0$ whereas $V_E(\chi)$ is given by a Taylor series starting from a
non-vanishing constant term $U_0$ as cosmological scales exit the horizon:
\begin{equation} \label{vechi}
K(\chi)=\frac{a_p}{\chi^p}+\cdots\,,\qquad V_E(\chi)=U_0(1-c\chi+\cdots)\,.
\end{equation}
In analogy with motion of a particle for $L=(1/2)m\dot{x}^2-V(x)$ and
$m\to\infty$, slow-roll inflation occurs for $\chi\to0$. We can calculate
the inflationary predictions using the usual slow-roll expressions (see
\sektion{calculate}) and
$\ud\sigma=\sqrt{K(\chi)}\ud\chi$, obtaining \cite{Galante:2014ifa}
\begin{equation} \label{Galante}
N_*\approx\frac{a_p\chi_*^{1-p}}{c(p-1)}\,,\quad n_s\approx1-\frac{p}{(p-1)N_*}\,,\quad
r\approx8\left(\frac{c^{p-2}a_p}{[(p-1)N_*]^p}\right)^{\frac{1}{p-1}}\,.
\end{equation}
For a standard thermal history after inflation, the current data
\cite{Ade:2015xua,Ade:2015lrj} favors $n_s\approx1-2/N_*$, which
corresponds to the case $p=2$. Note that for this case $r=8a_2/N_*^2$ does
not depend on $c$, which is to be expected since the kinetic term is
invariant under $\chi\to c\chi$. The Starobinsky model predictions given by
\eq{starpoint} correspond to $p=2$ and $a_2=3/2$.

Now consider inflation with $F(\phi)=1+\xi(\phi^2-v^2)$, so that $\chi\to0$
corresponds to $\xi(\phi^2-v^2)\gg1$. This implies that we can look for
Starobinsky-like solutions, with inflaton values above (below) the VEV if $\xi>0$ ($\xi<0$).
Using \eq{Zphi}, we obtain 
\begin{equation} \label{Zchi}
K(\chi)=\frac{3}{2\chi^2}+\frac{1}{4\xi\chi^2\left[1-\chi(1-\xi
v^2)\right]}\,.  
\end{equation} 
Note that this equation differs from eq.  (22) of ref. \cite{Galante:2014ifa}
due to the $\xi v^2$ term in  $F(\phi)$. As a consequence, the kinetic term
can remain positive after as well as during inflation for both signs of
$\xi$.

First, consider above VEV solutions satisfying $\phi^2-v^2\gg v^2$ as
cosmological scales exit the horizon. In this case
\begin{equation} \label{Zchi2}
K(\chi)\approx\frac{3\alpha}{2\chi^2}\,,
\text{ where }\alpha\equiv1+\frac{1}{6\xi}\approx \left\{
\arraycolsep=1.4pt\def\arraystretch{1.5}
\begin{array}{rl}
1 & \text{if } \xi \gg \frac16,\\
\frac{1}{6\xi} & \text{if }  \xi \ll \frac16.\\
\end{array} \right.
\end{equation}
Note that from \eq{Galante}, $\chi_*\approx a_2/(cN_*)$ with $a_2=3\alpha/2$,
so $\chi_*\ll1$ corresponds to $\xi\gg1/(4cN_*)$. Also, the assumption
$\phi^2-v^2\gg v^2$ corresponds to $\xi v^2\ll 2cN_*/(3\alpha)$.  When these
conditions are satisfied, the leading order inflationary predictions
coincide with those of the $\alpha$--attractor models
\cite{Ferrara:2013rsa,Kallosh:2013yoa,Galante:2014ifa}, namely,
\begin{equation}\label{alphaattractor}
n_s=1-\frac{2}{N_*}\,,\quad r=\frac{12\alpha}{N_*^2}\,.
\end{equation}

Second, let's assume that $|\phi^2-v^2|\ll v^2$ as cosmological scales exit
the horizon. Further assuming $6|\xi|v^2\gg|\phi^2-v^2|$, \eq{Zchi}
simplifies to $K(\chi)\approx3/(2\chi^2)$, that is, $p=2$ and
$a_2=3/2$. Therefore the Starobinsky model predictions given by
\eq{starpoint} are obtained for inflaton values both above and below the VEV whenever
these two assumptions are satisfied. Using $\chi_*\approx 3/(2cN_*)$, the two
assumptions correspond to $|\xi|v^2\gg(2cN_*)/3$ and $\xi^2v^2\gg cN_*/9$,
respectively. 

As long as these conditions are satisfied and $V_E(\chi)$ is given by
\eq{vechi}, the inflationary predictions will match the Starobinsky model
predictions up to leading order in $N_*$. From \eq{vechi} we obtain 
\begin{equation}
V_J(\phi)\approx
U_0\xi^2(\phi^2-v^2)^2\left(1+\frac{2-c}{\xi(\phi^2-v^2)}+\cdots\right)\,,
\end{equation}
which implies that regardless of the value of $c$ (as long as it is
$\ll1/\chi$), Starobinsky-like solutions are obtained if $V_J(\phi)$ is
approximately given by the double well potential for which $c=2$, as
cosmological scales exit the horizon. Therefore in terms of
$\varphi\equiv\phi-v$, the potentials satisfying \eq{vechi} can be written
as
\begin{equation}
V_J(\varphi)\propto\varphi^4\left(1+2\frac{v}{\varphi}\right)^2\,.
\end{equation} 
This again shows that the Starobinsky point given by \eq{starpoint} is
obtained for the $\varphi^4$
potential away from the minimum ($\varphi^2\gg v^2$), and the $\varphi^2$
potential near the minimum ($\varphi^2\ll v^2$). From
$\chi_*\approx3/(4N_*)$ (for $c=2$), the value of $\varphi$ as cosmological
scales exit the horizon is $\varphi_*^2\approx4N_*/(3\xi)$ and
$|\varphi_*|\approx2N_*/(3|\xi| v)$ for these two cases, respectively.

We can summarize the Starobinsky conditions as follows. The inflationary
predictions for $V_J(\phi)$ coincide with the Starobinsky model predictions
up to leading order in $N_*$ if:
\begin{description}[style=multiline,font=\normalfont]
\item[1.] The inflaton is above the VEV, $V_J(\phi)$ is quartic for $\phi^2\gg
v^2$, $\xi v^2\ll4N_*/3$ and $\xi\gg1/6$. (On the other hand, from \eq{Zchi2}, $r\approx2/(\xi
N_*^2)$ for $1/(8N_*)\ll\xi\ll1/6$.)
\item[2.] The potential is quadratic around the minimum for $\varphi^2\ll v^2$ and 
\begin{equation}\label{star2}
\xi^2v^2\gg \frac{2N_*}{9}  \text{ if }  |\xi|<\frac16\,,\quad
|\xi| v^2\gg \frac{4N_*}{3} \text{ if }  |\xi|>\frac16\,,\\
\end{equation}
\end{description}
where $\xi>0$ ($\xi<0$) for inflaton values above (below) the VEV. These
conditions satisfy the strong coupling limit \eq{strong}, so that a
plateau type Einstein frame potential is obtained during inflation in terms
of the canonical scalar field:
\begin{equation}
V_E(\sigma)\approx U_0\left(1-e^{-2\sigma/\sqrt6}\right)\,.
\end{equation}
Both 1. and 2. are  special cases of the strong coupling attractor
model \eq{3950}, with $n=2$ and $n=1$, respectively. The $n=1$ case is
discussed in ref. \cite{Kehagias:2013mya} and also belongs
to the class of the induced inflation models discussed in ref.
\cite{Giudice:2014toa}.

%%%%%%%%%%%%%%%%%%%%%%%%%%%%%%%%%%%%%%%%%%%%%%%%%%%%%%%%%%%%%%%%

\section{Double-well potential} \label{double}

In this section we analyze the prototypical symmetry breaking potential
\cite{Goldstone:1961eq}
\begin{equation}
V_J(\phi)=V_0\left[1-\left(\frac{\phi}{v}\right)^2\right]^2\,,
\end{equation}
referred to as the double-well potential, the Higgs potential or the Landau-Ginzburg potential. First, let us briefly review inflation with this potential for the minimal
coupling case, which was analyzed in several papers, see e.g. refs.
\cite{Vilenkin:1994pv,Linde:1994wt,Destri:2007pv,Kallosh:2007wm,Smith:2008pf,Rehman:2008qs,Martin:2013tda,Okada:2014lxa,Ashoorioon:2014jja}.
When inflation occurs near the minimum, in terms of $\varphi\equiv\phi-v$, the
potential is approximately quadratic: $V\approx(4V_0/v^2)\varphi^2$.
Since $\varphi_*^2\approx4N_*$ for quadratic inflation, the observable part of
inflation occurs near the minimum for $v^2\gg4N_*$. Then the quadratic
potential predictions of 
\begin{equation}\label{quadratic}
n_s\approx1-\frac{2}{N_*}\,,\quad r\approx\frac{8}{N_*}\,,\quad
\frac{\ud n_s}{\ud\ln k}\approx-\frac{2}{N^2}\,, 
\end{equation}
are obtained for inflation both below the VEV and above the VEV. 

\begin{figure}[!t]
\begin{center}
\scalebox{0.45}{\includegraphics{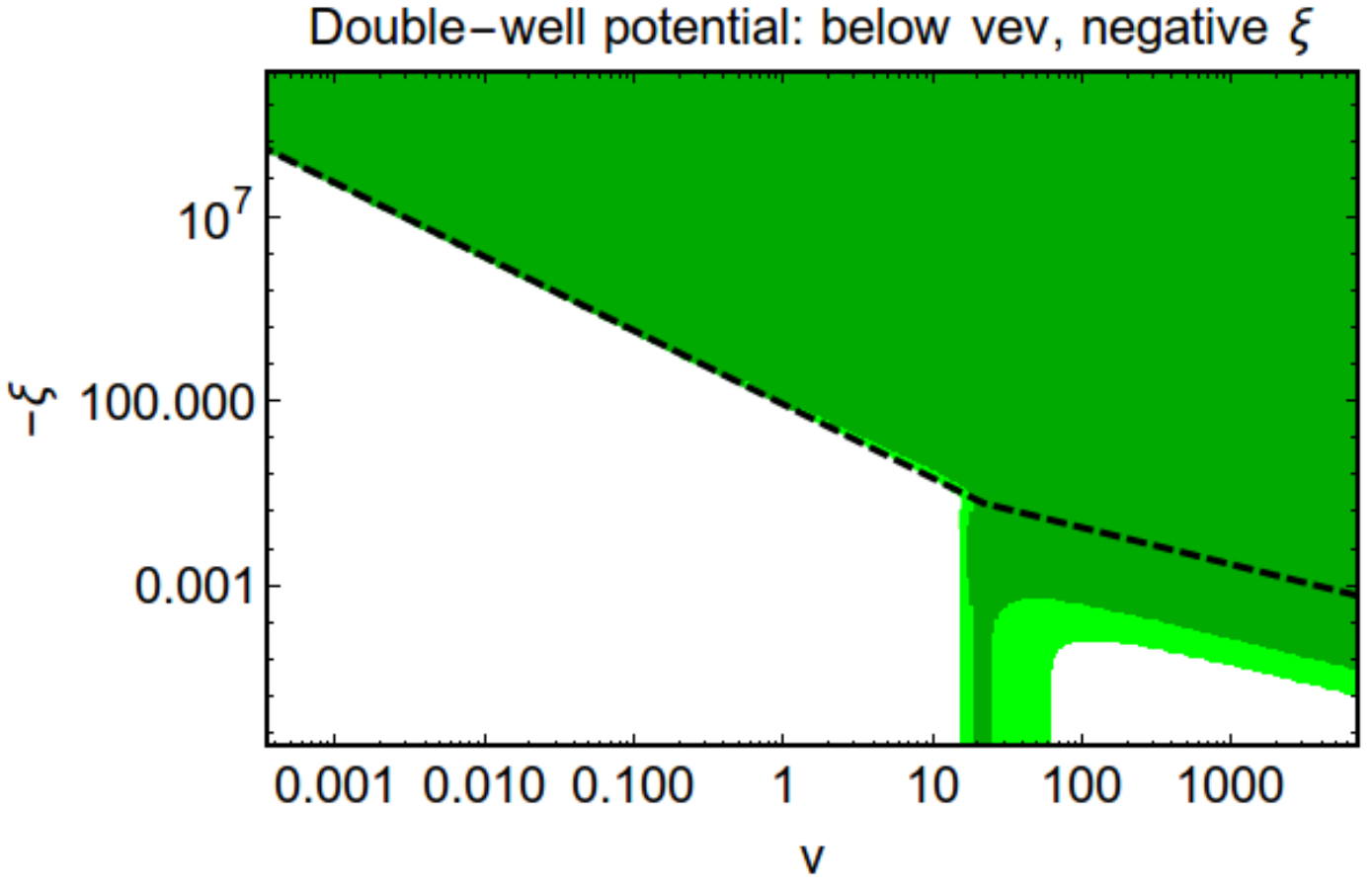}}\hspace{0.3cm}
\scalebox{0.45}{\includegraphics{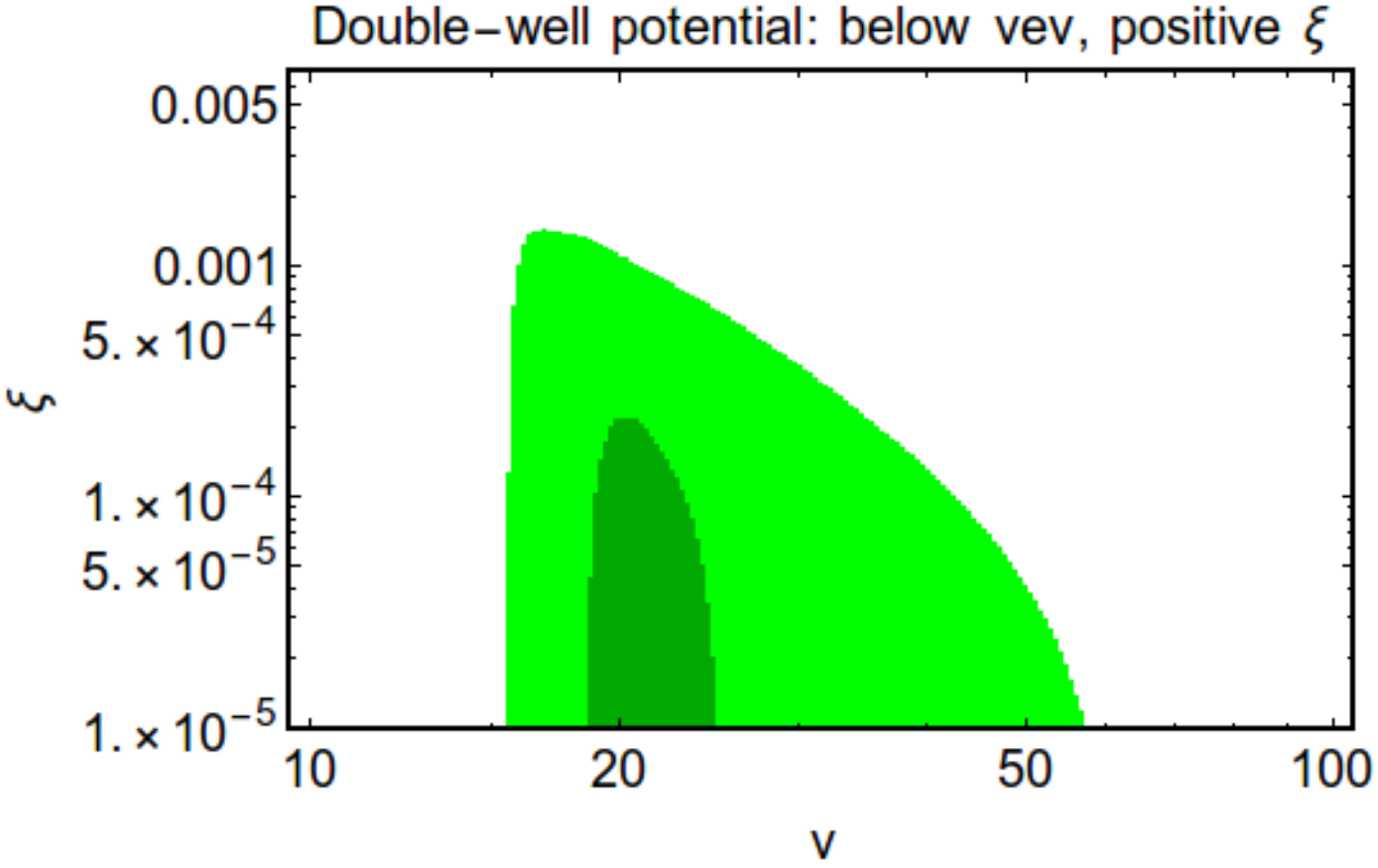}}
\\ \vspace{0.3cm}
\scalebox{0.45}{\includegraphics{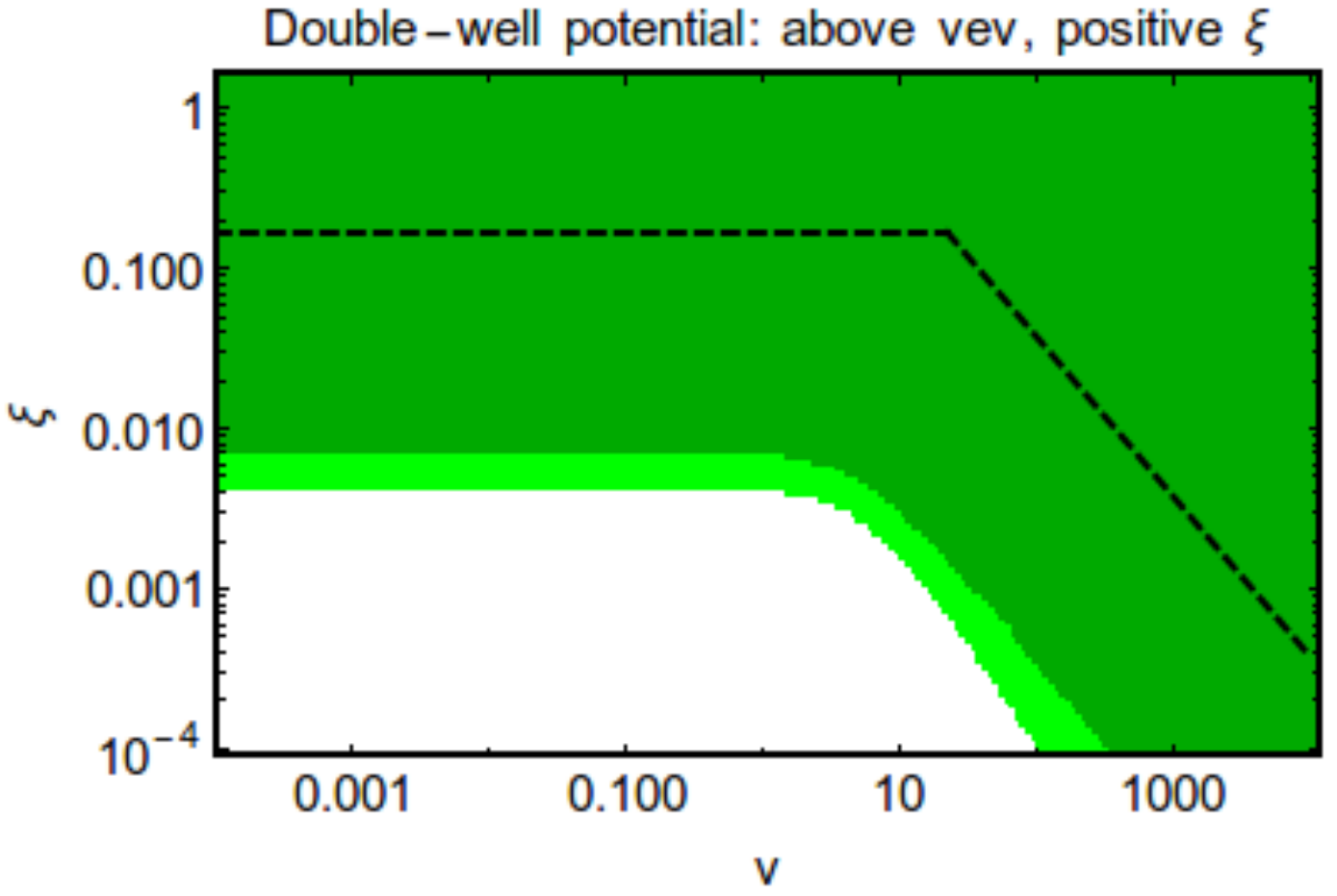}}
\end{center}\vspace{-0.5cm}
\caption{Light green  (green)  regions in the $v$-$\xi$ plane predict $n_s$ and $r$
values inside the 95\% (68\%) CL Planck contour (Planck TT+lowP+BKP+lensing+ext) \cite{Ade:2015xua}.  The Starobinsky
conditions \eq{star2} and \eq{higgsabove2} are satisfied above the dashed lines.}
  \label{higgs_vxi_figure}
\end{figure}

\begin{figure}[!t]
\centering
\includegraphics[angle=0, width=12cm]{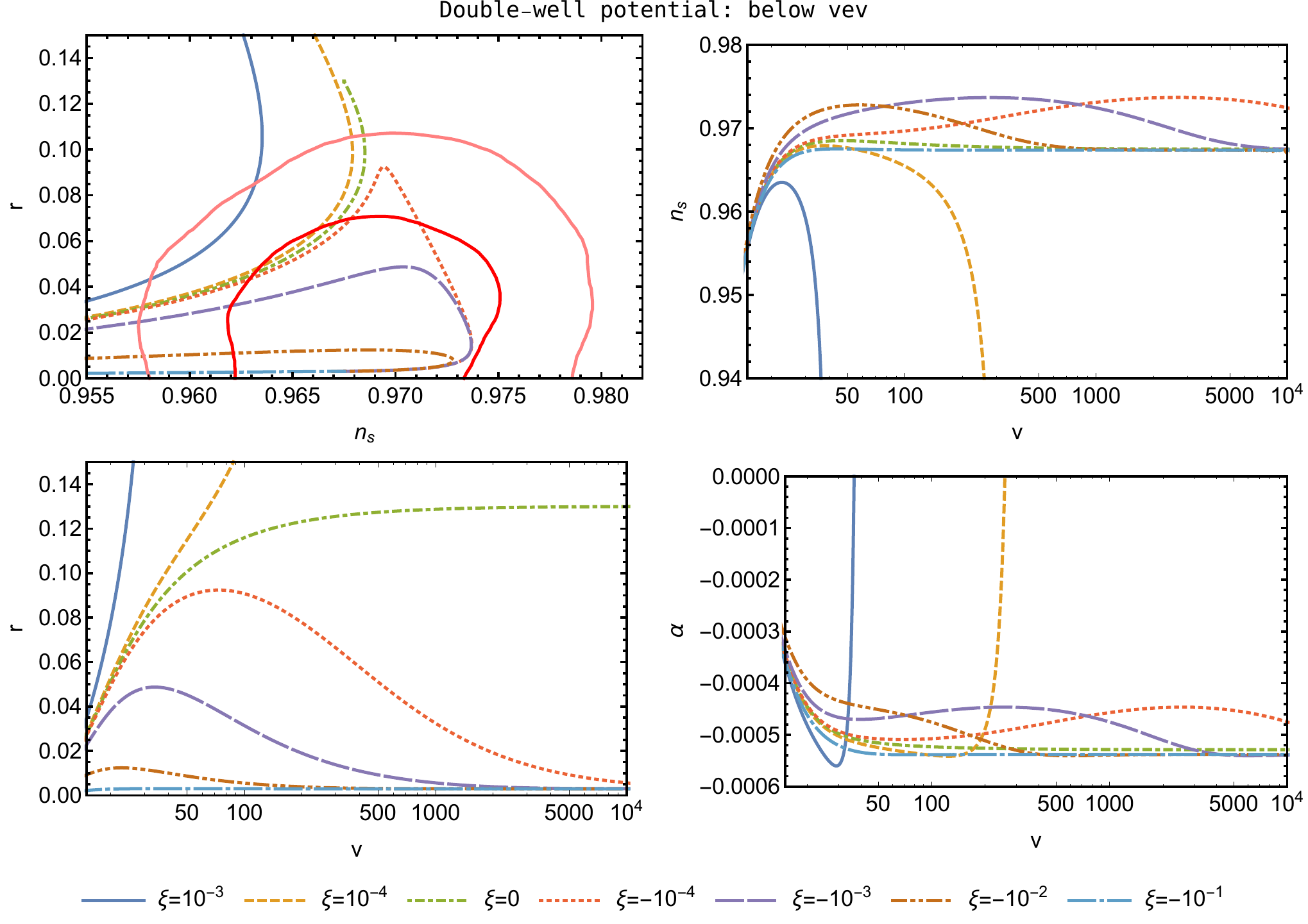}
\caption{Observational parameter values as functions of $v$ for selected $\xi$
values. The pink
(red) contour corresponds to the 95\% (68\%) CL contour
 given by the Planck collaboration (Planck TT+lowP+BKP+lensing+ext) \cite{Ade:2015xua}.}
  \label{higgs_below}
\end{figure}

For inflation above the VEV, if $v^2\ll4N_*$ then we have quartic inflation
with $n_s\approx1-3/N_*$ and $r\approx16/N_*$. The predictions interpolate between the
quadratic and quartic limits for $v^2\sim4N_*$, remaining out of the  95\%
CL Planck contour (Planck TT+lowP+BKP+lensing+ext)
\cite{Ade:2015xua} for all $v$.
Whereas for inflation below the VEV, if $v^2\ll4N_*$ then $\phi\ll v$ as
cosmological scales exit the horizon, so the potential is effectively of the new inflation
(small field or hilltop inflation) type
\begin{equation}
V(\phi)\approx V_0\left[1-2\left(\frac{\phi}{v}\right)^2\right]\,,
\end{equation}   
which implies a strongly red tilted spectrum $n_s\approx1-8/v^2$ with
suppressed $r$. As a result, although both the $v^2\ll4N_*$ and
$v^2\gg4N_*$ limits are ruled out, the $n_s$-$r$ values are in the 68\% CL
Planck contour (Planck TT+lowP+BKP+lensing+ext) \cite{Ade:2015xua} for a
narrow range around $v^2\sim4N_*$ (specifically, between $v=19$ and 25 for
the high-$N$ case), see figures \ref{higgs_vxi_figure} and
\ref{higgs_below}. Note that all the figures in this section are obtained
for the high-$N$ case, using the equations given in \sektion{calculate}. In
particular from \eq{efold2} we obtain:
\begin{equation}\nonumber
N_*=\frac18(1+6\xi)(\phi_*^2-\phi_e^2)+\frac{v^2}{4}\ln\frac{\phi_e}{\phi_*}+\frac34\ln\frac{1+\xi(\phi_e^2-v^2)}{1+\xi(\phi_*^2-v^2)}\,.
\end{equation}

From the discussions in \sektion{nonminimal}, we expect that we need
$\xi<0$ ($\xi>0$) to improve the fit to the observations for inflation
below (above) the VEV. Indeed, figures \ref{higgs_vxi_figure} and \ref{higgs_below}
show that for inflation below the VEV, the predictions move out of the 95\%
CL Planck contour for $\xi\gtrsim10^{-3}$. For inflation
above the VEV, the $\xi=0$ case interpolating between quadratic and quartic
inflation is already out of the Planck range, as is the $\xi<0$ case which
leads to an even redder spectrum and larger $r$.

It was discussed in \sektion{nonminimal} that if certain constraints on $v$
and $\xi$ values are satisfied, a potential quadratic near
its minimum or a potential quartic away from it correspond to
special cases of the strong coupling attractor model of
ref. \cite{Kallosh:2013tua}. The double-well potential satisfies both
conditions. The Einstein frame potential can be written as
\begin{equation}
V_E(\chi)=\frac{V_0}{\xi^2v^4}(1-2\chi+\chi^2)\,.
\end{equation}
Therefore the discussion in \sektion{star} is directly applicable. In
particular, \eq{star2} implies that for inflation below the VEV, as $|\xi|$
is increased for a given $v$, the predictions eventually approach the
Starobinsky point given by \eq{starpoint}. As can be seen from
\fig{higgs_vxi_figure}, this transition is rather abrupt for $v^2\ll8N_*$
and occurs near $|\xi|v^2=4N_*/3$. This means that if the Starobinsky point
is excluded by future observations, the entire parameter space $v^2\ll8N_*$
will be ruled out for below VEV inflation with this potential.

\begin{figure}[!t]
\centering
\includegraphics[angle=0, width=12cm]{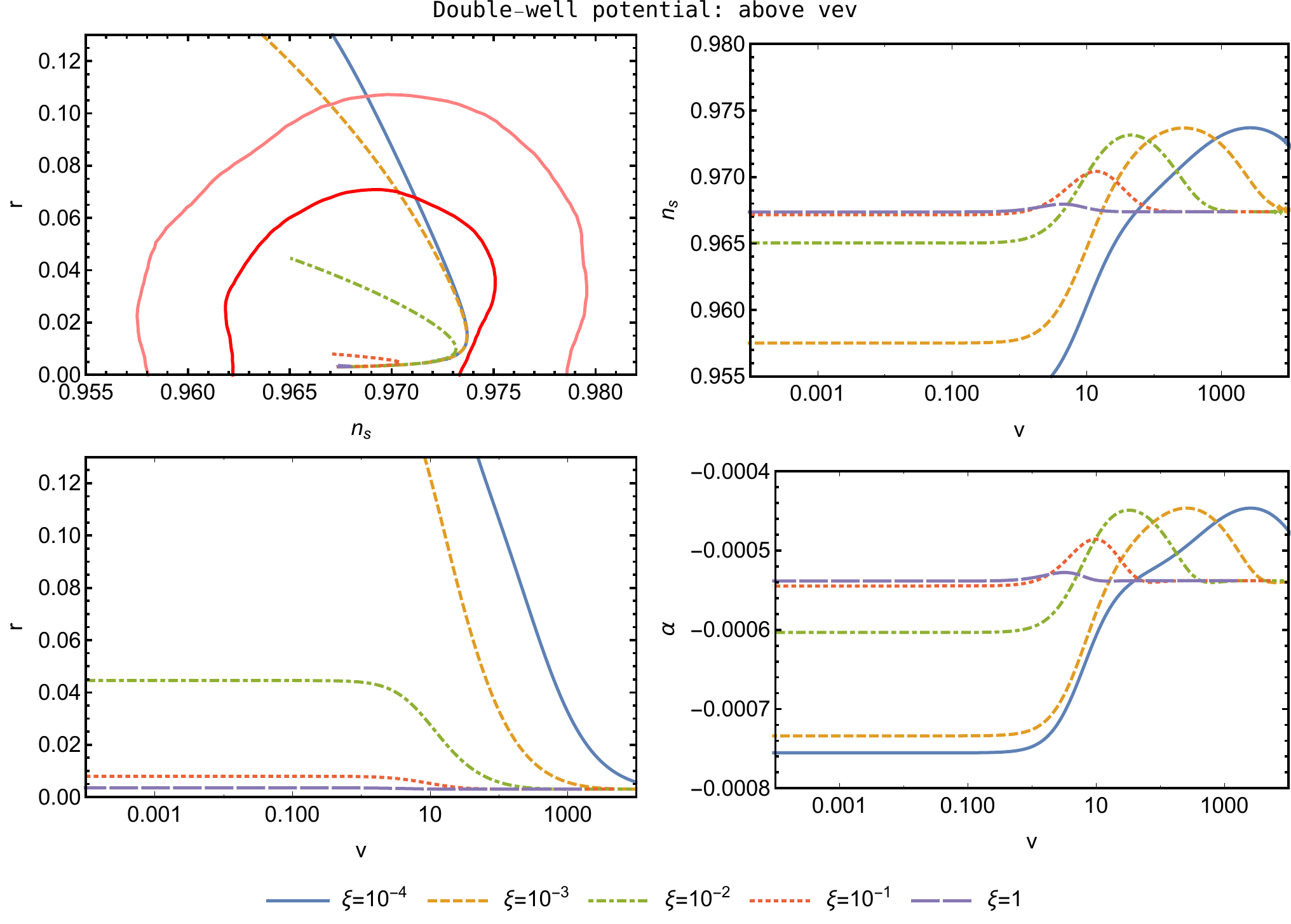}
\caption{Observational parameter values as functions of $v$ for selected $\xi$
values. The pink
(red) contour corresponds to the 95\% (68\%) CL contour
 given by the Planck collaboration (Planck TT+lowP+BKP+lensing+ext) \cite{Ade:2015xua}.}
  \label{higgs_above}
\end{figure}

For above VEV inflation with $\xi v^2\ll4N_*/3$ (which includes the induced
gravity case $\xi v^2=1$) and $\xi\gg1/(8 N_*)$, the inflationary
predictions are given by \eq{alphaattractor}.  Thus, the $n_s$-$r$ values
are in the 95\% (68\%) CL contours for $\xi\gtrsim0.004$ (0.007) for the
high-$N$ case, see figures \ref{higgs_vxi_figure} and \ref{higgs_above}.
Note that for very low reheat temperatures the $n_s$-$r$ values can move
out of the 68\% CL contour, see \fig{higgs_above_reheat}. For negligible
values of $v$, as is the case in standard model Higgs inflation
\cite{Bezrukov:2007ep}, the model is reduced to the non-minimally coupled
quartic inflation model. Our results in this limit agree with previous
results \cite{Fakir:1990eg,Salopek:1988qh,Okada:2010jf,Bezrukov:2013fca}.

Combining \eq{alphaattractor} with \eq{star2} we see that for above VEV inflation
the Starobinsky point is obtained when:
\begin{equation} \label{higgsabove2}
\xi^2v^2\gg \frac{2N_*}{9}  \text{ if }  0<\xi<\frac16\,,\quad
\forall v  \text{ if }  \xi\gg\frac16\,.
\end{equation}

In the induced gravity limit, using \eq{induced}, the Einstein frame potential can be written as
\begin{equation}
V_E(\sigma)=\frac{V_0}{\xi^2
v^4}\left(1-\exp\left[\frac{-2\sigma}{\sqrt{6\alpha}}\right]\right)^2\,,
\end{equation}
coinciding with the $\alpha$-$\beta$ model of refs.
\cite{Ferrara:2013rsa,Kallosh:2013yoa}. Thus the inflationary predictions
approach the quadratic potential predictions given by \eq{quadratic} for $\xi\ll1/(16N_*)$, and
\eq{alphaattractor} for larger $\xi$. The double-well potential in the induced
gravity limit was previously considered for inflation in refs.
\cite{Accetta:1985du,Lucchin:1985ip,Kaiser:1994vs,Cerioni:2009kn,Burns:2016ric}.
Ref. \cite{Kaiser:1994vs} also calculated $n_s$ for $\xi v^2$ values
between 0 and 1. Our results agree with ref. \cite{Kaiser:1994vs} for $\xi
v^2\ll1$ but not for the induced gravity limit $\xi v^2\to1$. For the
latter case our results agree with refs.
\cite{Cerioni:2009kn,Burns:2016ric}. 

Finally, ref. \cite{Linde:2011nh} analyzed the double-well potential with
non-minimal coupling in detail (see also ref. \cite{Tronconi:2017wps} for
the $\xi>0$ case).  The difference between our work and theirs is that we
take $F(\phi)=m^2+\xi\phi^2=1+\xi(\phi^2-v^2)$ as explained in
\sektion{nonminimal}, whereas they take $F(\phi)=1+\xi\phi^2$. As a
consequence, although the predictions on the $n_s$-$r$ plane look generally
similar, there are a few differences between our and their results.
Namely, for inflation below the VEV their predictions approach the
Starobinsky point given by \eq{starpoint} when $\xi v^2\to-1$, whereas our
predictions approach it when \eq{star2} is satisfied. For above VEV
inflation, their predictions approach \eq{starpoint} only for large values
of $\xi$, whereas our predictions approach it when \eq{higgsabove2} is
satisfied.

%%%%%%%%%%%%%%%%%%%%%%%%%%%%%%%%%%%%%%%%%%%%%%%%%%%%%%%%%%%%%%%%

\section{Small field inflation potentials}\label{small}

Consider new inflation type models where the
inflaton is below the VEV during inflation. For the double-well potential,
we see that consistency with observations require $|\xi|v^2\gtrsim4N_*/3$
so that very large $|\xi|$ values are needed for sub-Planckian values of
the VEV $v$. On the other hand, a potential which is flatter near the
origin could be compatible with observations even if $|\xi|\ll1/6$. As an
example we take a simple generalization of the double-well potential:
\begin{equation}\label{generalizedhiggs}
V_J(\phi)=V_0\left[1-\left(\frac{\phi}{v}\right)^p\right]^2\,,\quad(p>2)\,.
\end{equation}
For the weak coupling limit $|\xi|\ll1/6$ and $|\xi|v^2\ll1$, we have
$\phi\approx\sigma$. If $v^2\ll4N_*$
then $\sigma\ll v$ during inflation, and the Einstein frame potential can be written as
\begin{equation}\label{sfipotential}
V_E(\sigma)\approx
V_0\left[1-\left(\frac{\sigma}{\mu}\right)^p-2\xi\sigma^2\right]\,,
\end{equation}
where we have defined $\mu=v/2^{1/p}$.\footnote{This potential also arises in some supersymmetric new inflation models
\cite{Izawa:1996dv,Kawasaki:2003zv,Yamaguchi:2004tn,Senoguz:2004ky}
and was analyzed in refs. \cite{Boyanovsky:2007ry,Destri:2009wn}.}

\begin{figure}[!t]
\centering
\includegraphics[angle=0, width=14cm]{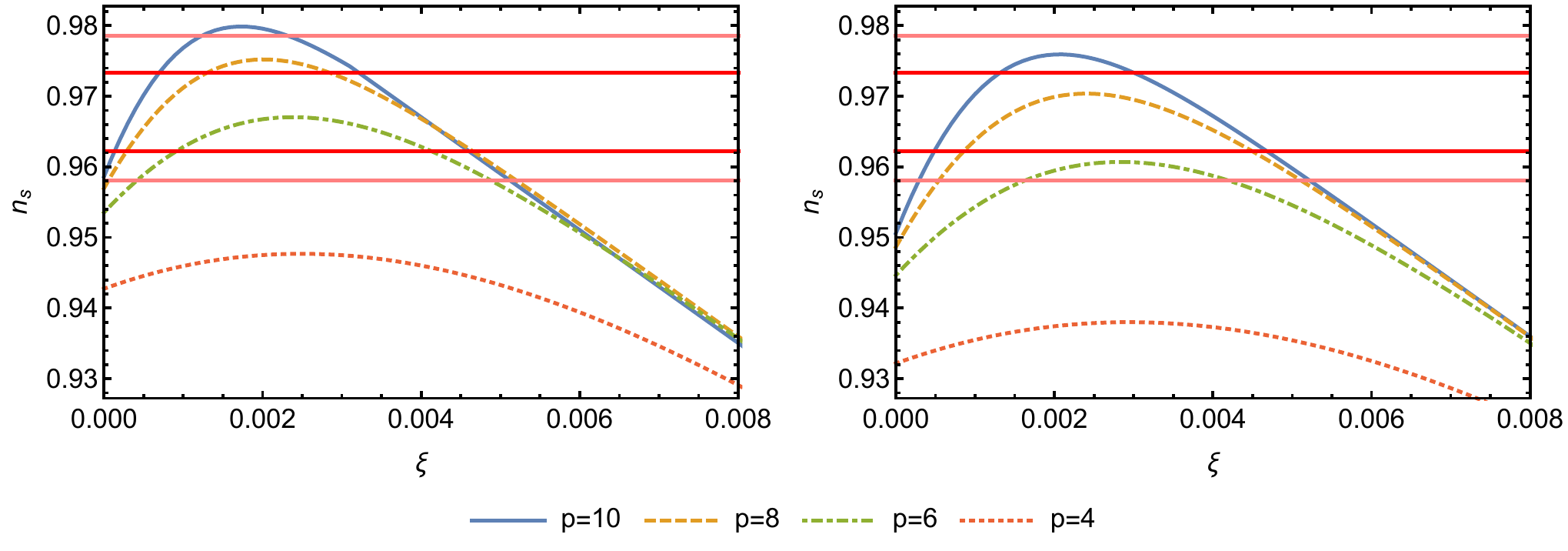}
\caption{$n_s$ values as functions of $\xi$ for $\mu=0.01$ and selected $p$
values. Left panel: high-$N$ scenario, right panel: low-$N$ scenario (see
\sektion{nonminimal}). The pink
(red) line corresponds to the 95\% (68\%) CL contour
 given by the Planck collaboration (Planck TT+lowP+BKP+lensing+ext) \cite{Ade:2015xua}.}
  \label{sfi_figure}
\end{figure}

For $\xi=0$, this small field inflation potential (also called hilltop
potential) appears often in the literature, see for example refs.
\cite{Boubekeur:2005zm,Lyth:2009zz,Martin:2013tda} and references
therein. Using eqs. \ref{slowroll1}, \ref{nsralpha1} and \ref{efold1}, we obtain
\begin{equation}\label{nsrminimal}
n_s\approx1-\frac{(p-1)2}{(p-2)N_*}\,,\quad
r\approx128\left(\frac{16\mu^{2p}}{p^2[4(p-2)N_*]^{2p-2}}\right)^\frac{1}{p-2}\,,\end{equation}
which shows that $r$ is suppressed and $n_s$ tends to be smaller than the range
favored by observations. To be more specific, let's consider the most
optimistic high-$N$ case, where using eqs. \ref{efolds}, \ref{nstarandr}, and
$\rho_*\approx\rho_e$, we have
\begin{equation}
N_*\approx64.7+\frac14\ln\frac{3\pi^2\Delta_\mathcal{R}^2 r}{2}\,.
\end{equation}
Note that the energy scale during inflation is lower for lower $\mu$ values,
which correspond to lower $r$, $N_*$ and therefore $n_s$ values as well.
For $\mu=0.01$, $n_s$ can be inside the  95\% CL contour
 given by the Planck collaboration (Planck TT+lowP+BKP+lensing+ext)
\cite{Ade:2015xua} only for $p\ge10$, see \fig{sfi_figure}. If
$\mu\lesssim6\times10^{-8}$, $n_s$ is outside the 95\% CL contour for any $p$
value.

Repeating the calculation for the potential given by \eq{sfipotential}, we obtain
\begin{equation}\label{sfinsr}
n_s\approx1+\frac{8(p-1)\xi}{1-e^{4(p-2)\xi N_*}}-8\xi\,,\quad
r\approx\frac{128\xi^2\mu^2(4\xi\mu^2/p)^{2/(p-2)}e^{8(p-2)\xi N_*}}{
\left(e^{4(p-2)\xi N_*}-1\right)^{2(p-1)/(p-2)}}\,.
\end{equation}
These expressions are in excellent agreement with the numerical results
given in \fig{sfi_figure}, which were calculated using the Jordan frame
potential given by \eq{generalizedhiggs}. They show that $n_s$ values increase and
the fit to observational data improves provided $\xi\sim1/[4(p-2)N_*]$. In
particular, $n_s$ can be inside the 95\% CL contour for much smaller VEVs,
namely for $\mu\gtrsim2\times10^{-9}$, $2\times10^{-17}$, $7\times10^{-23}$
for $p=6$, 8, 10 respectively.
   
%%%%%%%%%%%%%%%%%%%%%%%%%%%%%%%%%%%%%%%%%%%%%%%%%%%%%%%%%%%%%%%%

\section{Coleman-Weinberg potential} \label{cw}

Symmetry breaking due to the Coleman-Weinberg mechanism
\cite{Coleman:1973jx} was associated with inflation since the early
eighties when the first new inflation models were proposed
\cite{Linde:1981mu,Albrecht:1982wi,Shafi:1983bd}. In these models the
effective potential can be written as \cite{Albrecht:1984qt,Linde:2005ht}
\begin{equation} \label{potpot}
                  V_J(\phi)= A \phi^4 \left[\ln\left( \frac{\phi}{v}\right)
-\frac{1}{4}\right] + \frac{A v^4}{4}\,.
\end{equation}
For a minimally coupled scalar, the inflationary predictions of this
potential were analyzed in ref. \cite{Shafi:2006cs} (see also refs.
\cite{Smith:2008pf,Rehman:2008qs,Martin:2013tda,Barenboim:2013wra,Okada:2014lxa,Kannike:2014mia,Senoguz:2015lba}).
They are generally similar to the predictions of the double-well potential:
Again, for $v^2\gg4N_*$ inflation occurs around the quadratic minimum
$V\approx2Av^2\varphi^2$, leading to \eq{quadratic}. For inflation above
the VEV, the predictions again interpolate between the quadratic and
quartic limits, remaining out of the 95\% CL Planck contour. Whereas for
inflation below the VEV, if $v^2\ll4N_*$ then $\phi\ll v$ as cosmological
scales exit the horizon, so the potential is effectively of the new
inflation (small field or hilltop inflation) type
\begin{equation}\label{approxpot}
V(\phi)\approx V_0\left[1-\left(\frac{\phi}{\mu}\right)^4\right]\,,
\end{equation} 
which predicts $n_s\approx1-3/N_*$, $\alpha\approx-3/N_*^2$ and a tiny $r$
as given by \eq{nsrminimal}. Comparing with \eq{potpot}, the parameter
$\mu$ in \eq{approxpot} is given by
\begin{equation}
\mu^4\approx-\frac{v^4}{4}\left(\ln\frac{\phi_*}{v}-\frac14\right)^{-1}\,,
\end{equation}
where using  \eq{efold1} we obtain
\begin{equation}
\left(\frac{\phi_*}{v}\right)^2\approx-\frac{v^2}{16N_*}\left[W_{-1}\left(-\frac{v^2}{16N_*}\right)\right]^{-1}\,.
\end{equation}
Here, $W_{-1}$ is a branch of the Lambert function satisfying $W(z)e^{
W(z)}=z$.

\begin{figure}[!t]
\begin{center}
\scalebox{0.45}{\includegraphics{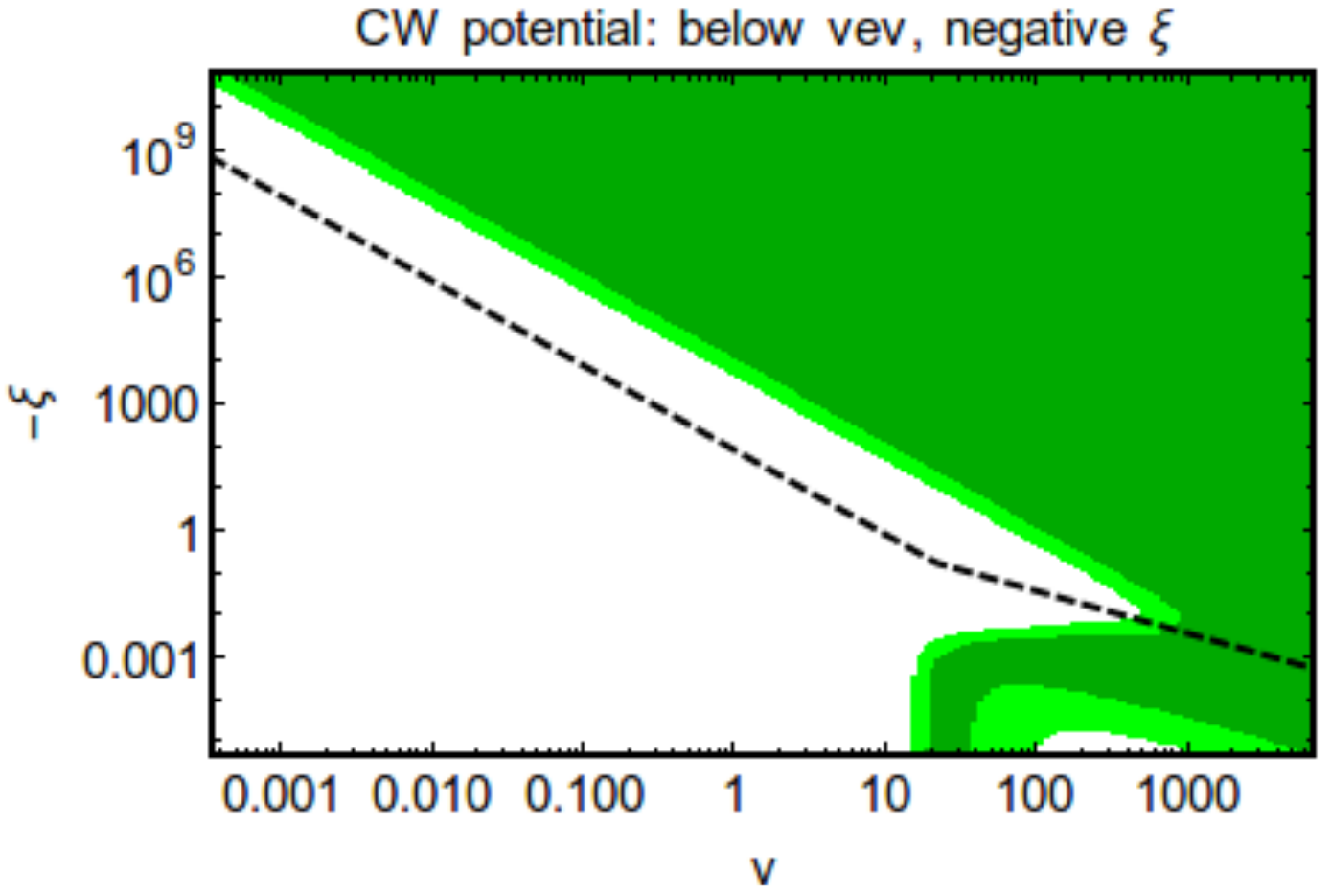}}\hspace{0.3cm}
\scalebox{0.42}{\includegraphics{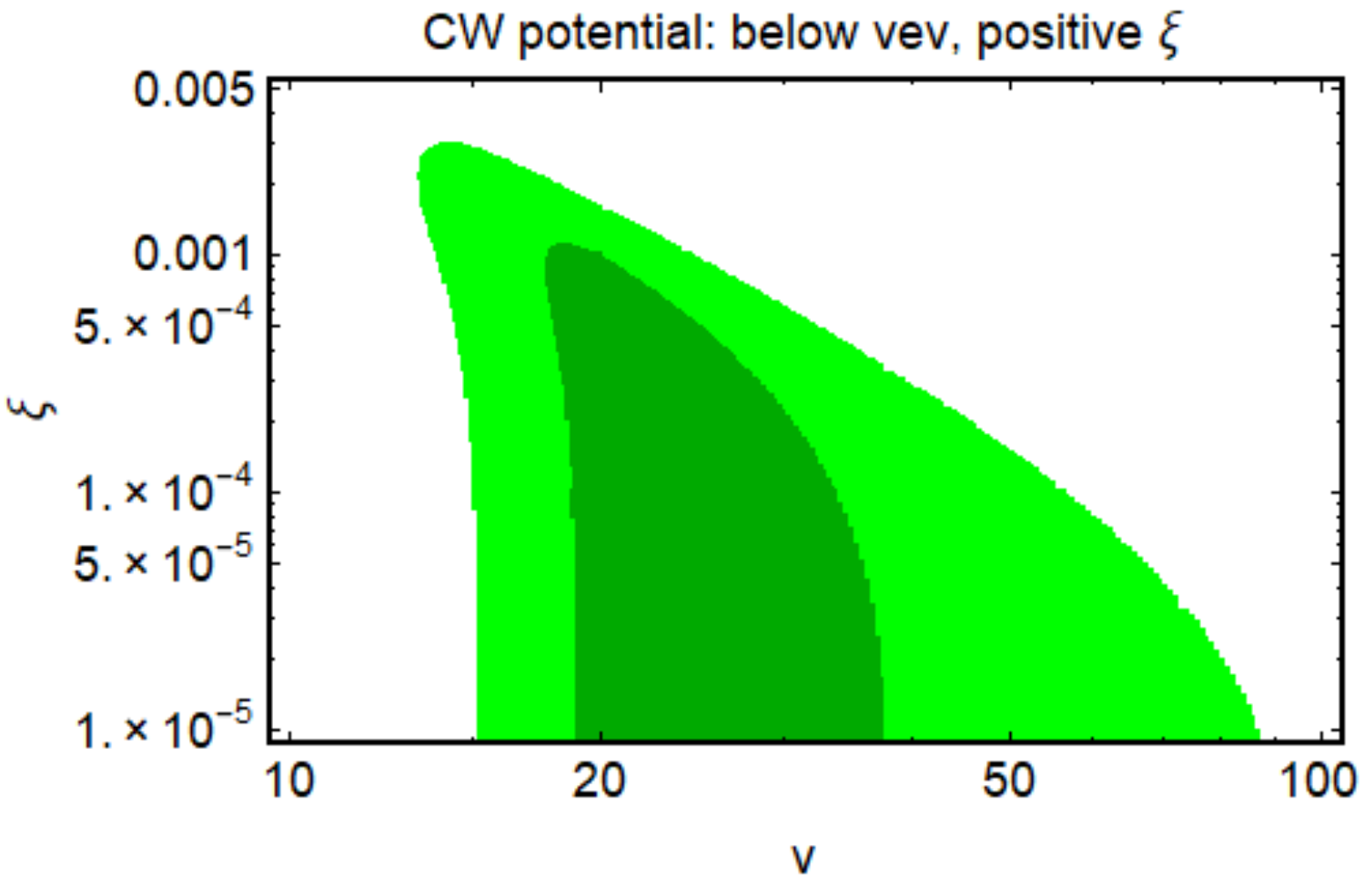}}
\\ \vspace{0.3cm}
\scalebox{0.54}{\includegraphics{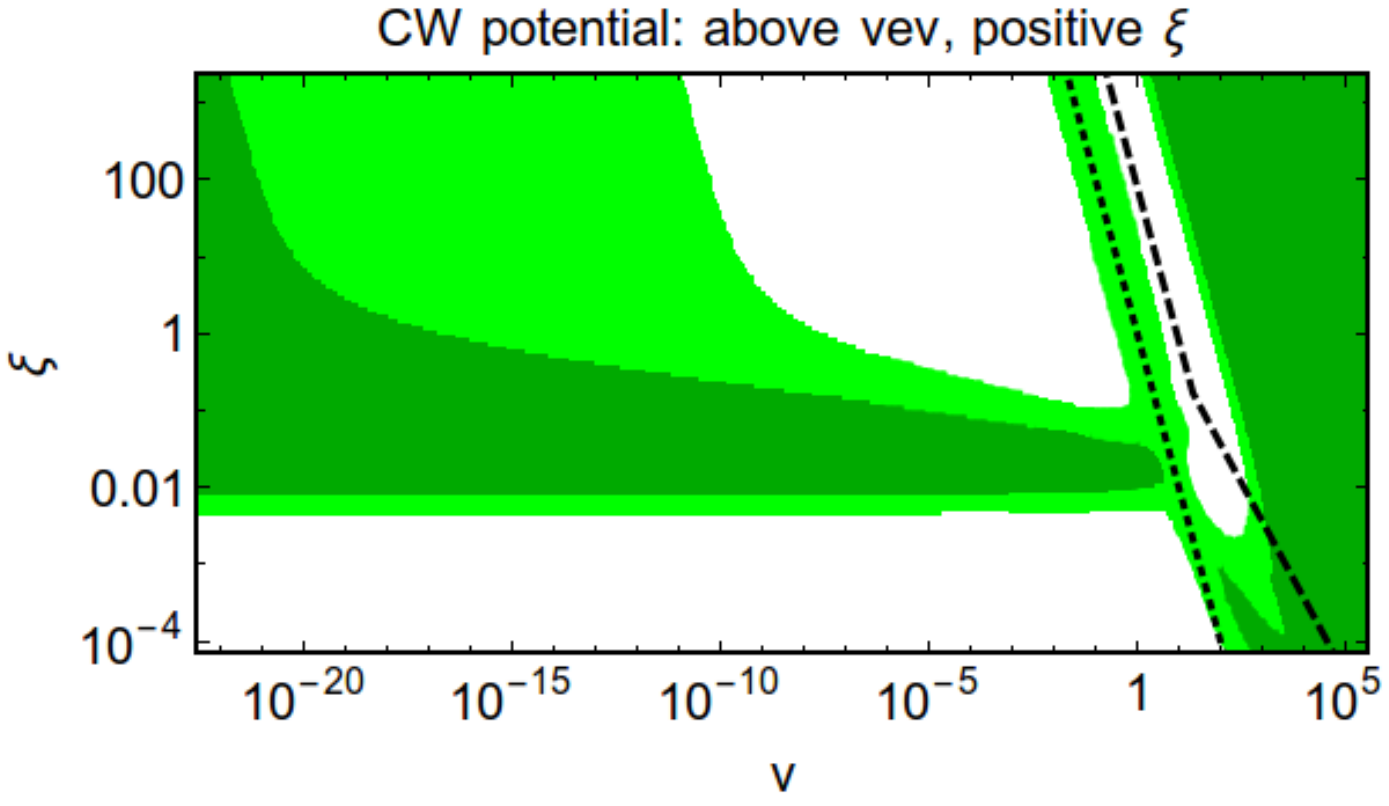}}
\end{center}\vspace{-0.5cm}
\caption{Light green (green) regions in the $v$-$\xi$ plane predict $n_s$ and $r$
values inside the 95\% (68\%) CL Planck contour (Planck TT+lowP+BKP+lensing+ext) \cite{Ade:2015xua}. The Starobinsky
conditions \eq{star2} are satisfied above the dashed lines. The
dotted line corresponds to $\xi v^2=1$. }
  \label{cw_vxi_figure}
\end{figure}

\begin{figure}[!b]
\centering
\includegraphics[angle=0, width=12cm]{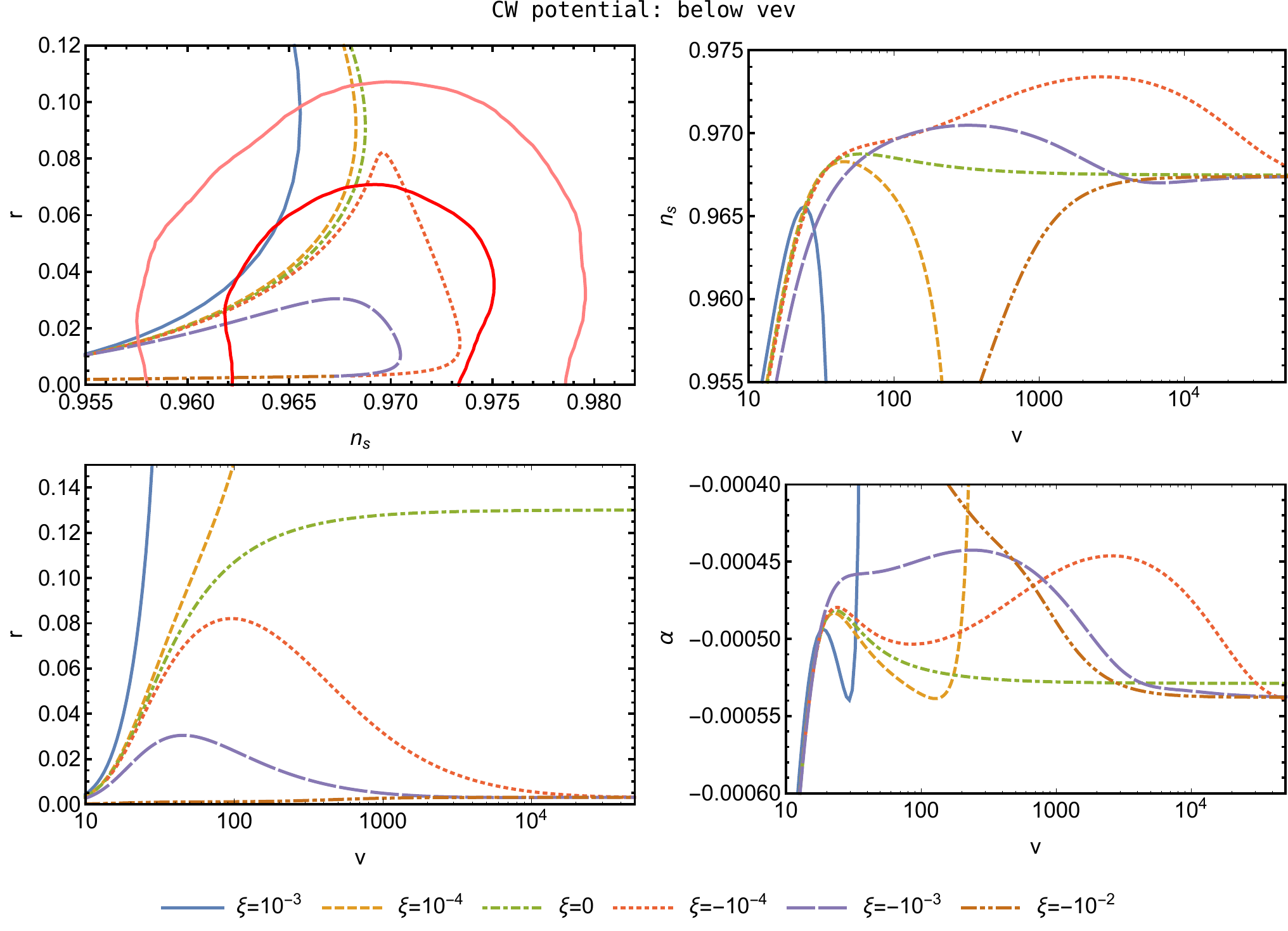}
\caption{Observational parameter values as functions of $v$ for selected $\xi$
values. The pink
(red) contour corresponds to the 95\% (68\%) CL contour
 given by the Planck collaboration (Planck TT+lowP+BKP+lensing+ext) \cite{Ade:2015xua}.}
  \label{cw_below}
\end{figure}

Similarly to the double-well potential, although both the $v^2\ll4N_*$ and
$v^2\gg4N_*$ limits are ruled out for inflation below the VEV, 
the $n_s$-$r$ values are in the 68\% CL
Planck contour (Planck TT+lowP+BKP+lensing+ext) \cite{Ade:2015xua} for a
narrow range around $v^2\sim4N_*$ (specifically, between $v=20$ and 38 for
the high-$N$ case), see figures \ref{cw_vxi_figure} and
\ref{cw_below}. Note that all the figures in this section are also obtained
for the high-$N$ case.

From \fig{cw_vxi_figure} we see that below VEV inflation predictions are
incompatible with the observational data for $|\xi|\ll1/6$ and
$v^2\lesssim1$. This is expected from the discussion in \sektion{small}
since under these conditions the Coleman-Weinberg potential is
approximately given by \eq{sfipotential} with $p=4$ during inflation.  As
is clear from \fig{sfi_figure} and \eq{sfinsr}, a small non-minimal
coupling cannot bring $n_s$ into agreement with observations. For instance
$n_s$ remains $\le0.945$ for $v=0.01$ even under the most favorable instant
reheating assumption. This result agrees with refs.
\cite{Iso:2014gka,Kaneta:2017lnj}, but disagrees with ref.
\cite{Panotopoulos:2014hwa} where the quartic term in \eq{sfipotential} is
erroneously neglected.

The effect of the non-minimal coupling on the predictions of inflation below 
the VEV is similar for Coleman-Weinberg potential to the double-well
potential. Both potentials are quadratic near their minima, so that the
Starobinsky point given by \eq{starpoint} is obtained when \eq{star2}
holds. Numerically, approaching the Starobinsky point requires even larger
$|\xi|v^2$ for the Coleman-Weinberg potential as can be seen from
\fig{cw_vxi_figure}. The predictions move out of the 95\%
CL Planck contour for $\xi\gtrsim3\times10^{-3}$.

\begin{figure}[!t]
\centering
\includegraphics[angle=0, width=12cm]{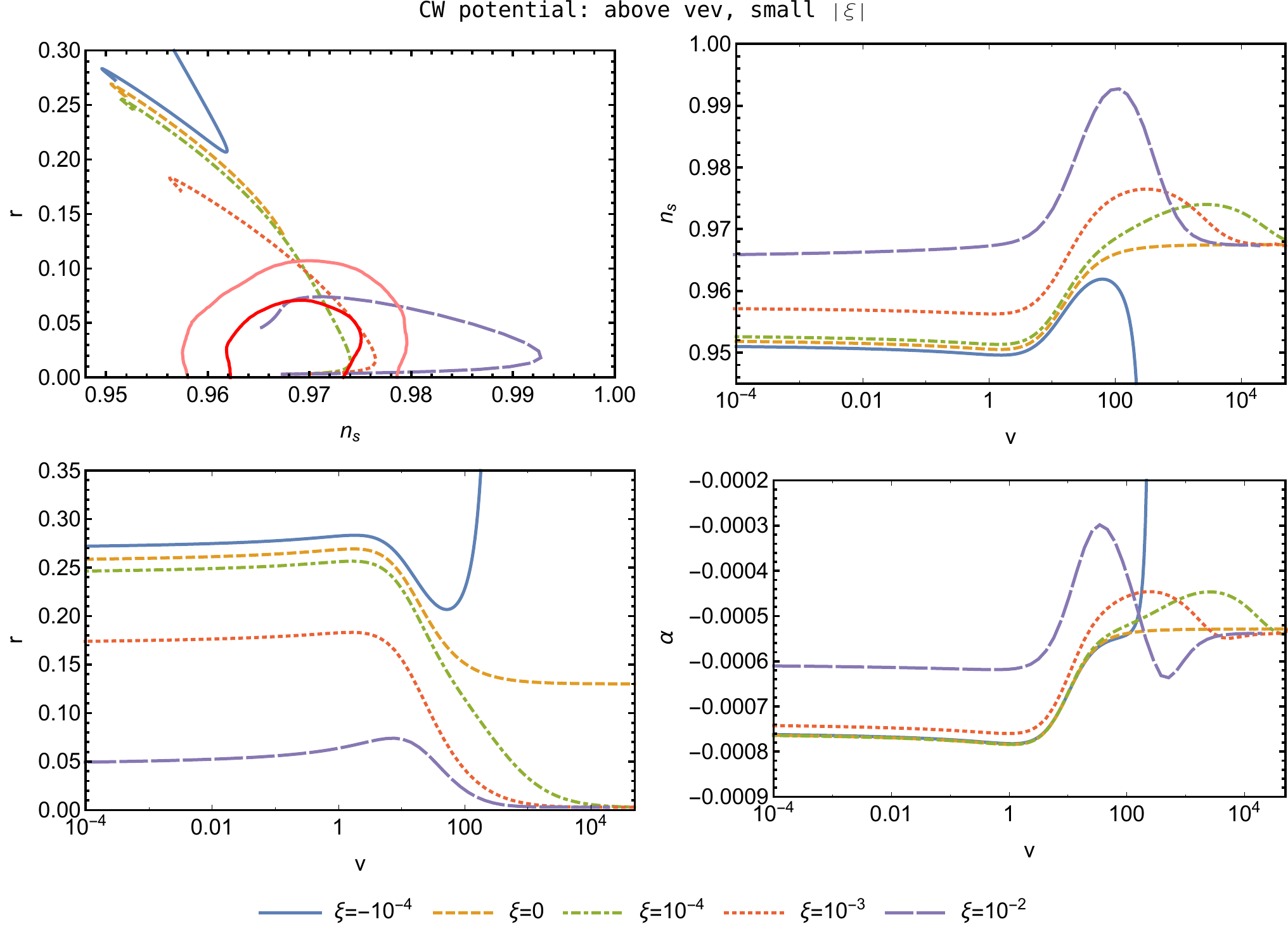}
\caption{Observational parameter values as functions of $v$ for selected $\xi$
values. The pink
(red) contour corresponds to the 95\% (68\%) CL contour
 given by the Planck collaboration (Planck TT+lowP+BKP+lensing+ext) \cite{Ade:2015xua}.}
  \label{cw_above1}
\end{figure}

\begin{figure}[!t]
\centering
\includegraphics[angle=0, width=12cm]{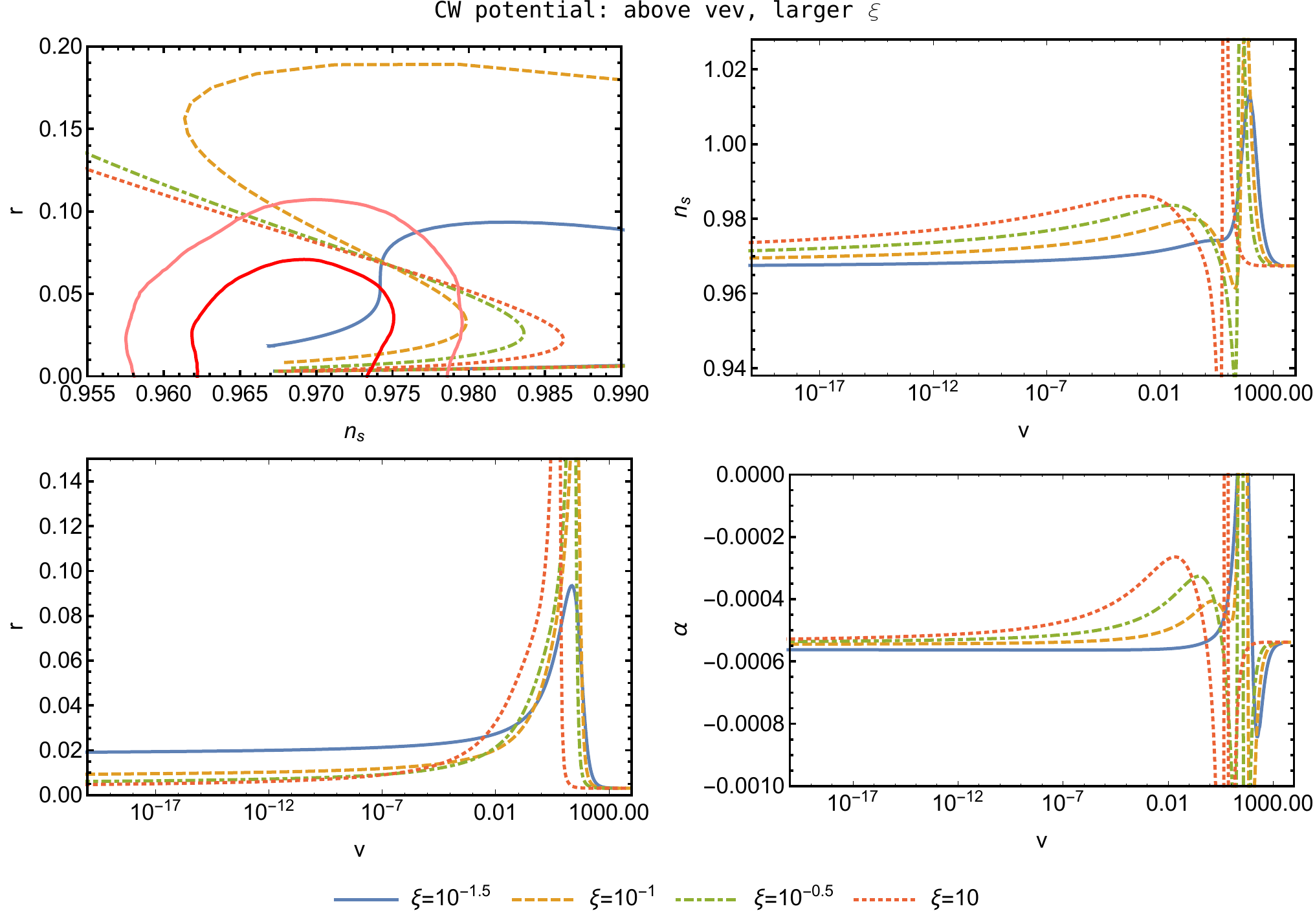}
\caption{Observational parameter values as functions of $v$ for selected $\xi$
values. The pink
(red) contour corresponds to the 95\% (68\%) CL contour
 given by the Planck collaboration (Planck TT+lowP+BKP+lensing+ext) \cite{Ade:2015xua}.}
  \label{cw_above2}
\end{figure}

For inflation above the VEV, the $\xi=0$ case interpolating between
quadratic and quartic inflation is already out of the Planck range, as is
the $\xi<0$ case which leads to an even redder spectrum and larger $r$ (see
\fig{cw_above1}). The $\xi>0$ case, analyzed in ref.
\cite{Marzola:2015xbh}, is more subtle since the
Coleman-Weinberg potential in the Jordan frame is not simply quartic away
from the minimum but also contains a logarithmic factor. There is similarly
a logarithmic factor in the Einstein frame potential written in terms of
$\chi$:
\begin{equation} \label{vechi3}
V_E(\chi)\approx\frac{A}{2\xi^2}\left[-\ln(\xi
v^2\chi)\right](1-2\chi)\,.
\end{equation}
Thus, the predictions of \eq{alphaattractor} are approached only when the
logarithmic factor can be treated as constant, that is, when the
contribution from its derivative can be neglected. Taking derivative of
\eq{vechi3} and using $\chi_*\approx 3\alpha/(4N_*)$ (see \sektion{star}), we
find that this requires
\begin{equation}
\xi v^2\ll\frac{4N_*}{3\alpha}\exp\left[-\frac{2N_*}{3\alpha}\right]\text{ for } 
\xi\gg\frac{1}{8N_*}\,.
\end{equation}
Numerically, $n_s$-$r$ values are in the 95\% (68\%) CL contours for
$\xi\gtrsim0.005$ (0.008), assuming the high-$N$ case and $v\ll1$, see
figures \ref{cw_vxi_figure} and \ref{cw_above1}. The Starobinsky point
given by \eq{starpoint} is obtained both for $\xi v^2\gg4N_*/3$ and for
$\xi\gg1/6$ with extremely small values of $\xi v^2$, whereas the
predictions move out of the observationally favored region in the $n_s$-$r$
plane as $\xi v^2$ approaches $4N_*/3$ (see \fig{cw_above2}).

Using \eq{induced} we can write the Einstein frame potential in
the induced gravity limit ($\xi v^2=1$) as follows \cite{Kannike:2015kda}:
\begin{equation} 
V(\sigma)=\frac{A}{4\xi^2}\left(4\sqrt{\frac{\xi}{1+6\xi}}\sigma
+\exp\left(-4\sqrt{\frac{\xi}{1+6\xi}}\sigma\right)-1\right)\,,
\end{equation}
where $\sigma>0$ ($\sigma<0$) during inflation above (below) the VEV.
Analysis of this potential
\cite{Cerioni:2009kn,Kannike:2015kda,Karam:2017zno} shows that inflation
below the VEV is not compatible with the current observational data,
whereas above VEV inflation predictions interpolate between the linear
potential and quadratic potential predictions for $v^2\ll2N_*$ and
$v^2\gg2N_*$, respectively. The linear potential predictions 
\begin{equation}\label{linear}
n_s\approx1-\frac{3}{2N_*}\,,\quad r\approx\frac{4}{N_*}\,,\quad
\frac{\ud n_s}{\ud\ln k}\approx-\frac{3}{2N^2}\,, 
\end{equation}
are in the Planck \%95 CL contour, which explains the light green region
around the dotted line in \fig{cw_vxi_figure}.

%%%%%%%%%%%%%%%%%%%%%%%%%%%%%%%%%%%%%%%%%%%%%%%%%%%%%%%%%%%%%%%%

\section{Conclusion} \label{conclude}

In this work we discussed the inflationary predictions of models with the
Lagrangian
\begin{equation} 
\frac{\mathcal{L}_J}{\sqrt{-g}}=\frac12(m^2+\xi\phi^2)R-\frac12g^{\mu\nu}\partial_{\mu}\phi\partial_{\nu}\phi-V_J(\phi)\,,
\end{equation}
where the inflaton $\phi$ has a non-zero VEV $v$ after inflation, and
$m^2=1-\xi v^2$. In terms of the redefined field $\varphi\equiv\phi-v$,
the non-minimal coupling in the Lagrangian includes a linear term in
$\varphi$ as well as a quadratic term. This leads to an attractor behaviour
where the predictions approach the Starobinsky model predictions, not just
for the well-known non-minimally coupled quartic potential case but also when
the inflaton is near the minimum ($\varphi^2\ll v^2$) as cosmological
scales exit the horizon.

After discussing the conditions under which Starobinsky-like behaviour is
obtained in \sektion{nonminimal}, we analyze two prototypical
symmetry breaking potentials: the double-well potential in \sektion{double}
and the Coleman-Weinberg potential in \sektion{cw}. For each potential, we
display the regions in the $v$-$\xi$ plane for which the spectral index
$n_s$ and the tensor-to-scalar ratio $r$ values are compatible with the
current observations.  

If $\xi>0$ ($\xi<0$) for inflation above (below) the VEV, large portions of
the $v$-$\xi$ plane lead to predictions compatible with the current
constraints on $n_s$ and $r$, see figures \ref{higgs_vxi_figure} and
\ref{cw_vxi_figure}. Most of these portions lead to predictions approaching
the Starobinsky model predictions, so the allowed parameter space would
shrink drastically if future observations rule out the Starobinsky model.
In particular, if the upper bound on $r$ becomes $<0.002$, both the
double-well and Coleman-Weinberg potentials would be ruled out as
inflationary models, for any value of $v$ and $\xi$.\footnote{All the
results in this article is based on the metric formulation of gravity. If
different approaches such as the Palatini formulation is used $r$ could be
much smaller \cite{Bauer:2008zj,Jarv:2017azx}.}

Although we have displayed the inflationary predictions for a wide range of
$v$ and $\xi$ values, it is questionable whether the entire range can be
theoretically justified. In particular $v\gtrsim 1$ can be difficult to
realize starting from a fundamental theory \cite{Baumann:2014nda}. The
value of $\xi$ is ambiguous unless the inflationary part of the Lagrangian
is embedded in a specific theory (see e.g.
\cite{Muta:1991mw,Faraoni:2004pi}).  However, for the well-known
non-minimally coupled quartic potential solution, expanding the action
around the vacuum reveals a cut-off scale $\Lambda= 1/\xi$
\cite{Burgess:2009ea,Barbon:2009ya,Hertzberg:2010dc}, and requiring this to
be higher than the energy scale during inflation corresponds to
$\xi\lesssim300$ using \eq{nstarandr} and \eq{starpoint}. On the other
hand, ref. \cite{Bezrukov:2010jz} has emphasized that the cut-off scale
depends on the background value of the field and can remain above the
relevant energy scales during and after inflation. In any case consistency
with observations only require $\xi\gtrsim0.005$ for this solution.

Starobinsky-like inflationary predictions also arise when \eq{star2} is
satisfied. The observable part of inflation then occurs near the minimum
where the potential is $\propto\varphi^2$, and $F(\varphi)\approx1+2\xi
v\varphi$. This is another special case of the strong coupling attractor
model \cite{Kallosh:2013tua}. Interestingly, even though the
Starobinsky-like regime corresponds to $|\xi|v\gg 1$ (with $\xi>0$ and
$\xi<0$ for inflation above and below the VEV, respectively), the cut-off
remains at the Planck scale \cite{Kehagias:2013mya,Giudice:2014toa}. Thus,
although consistency with observations require large values of $|\xi|$ for
sub-Planckian VEVs, no perturbative unitarity violation is expected at
scale $1/\xi$ around the vacuum, unlike the non-minimally coupled quartic
potential case.\footnote{The cut-off scale around the vacuum changes
when \eq{star2} is satisfied since expanding the Einstein frame Lagrangian
for small values of $\varphi$, the leading order kinetic term is no longer
canonically normalized but instead given by $1+6\xi^2 v^2\gg1$.} 

Finally, in \sektion{small}, we briefly considered a higher order version
of the double-well potential, which for sub-Planckian VEVs corresponds to a
small field (hilltop) inflation potential, with an additional quadratic
term coming from the non-minimal coupling. Unlike the two above-mentioned
potentials, for this potential inflation below a sub-Planckian VEV can be
compatible with observations for a positive $\xi\lesssim0.005$, and a tiny
$r$ is predicted. All the considered potentials predict a running of the
spectral index that is too small to be observed in the near future, with
$\ud n_s/\ud\ln k$ typically around $-2/N_*^2$.
   
%%%%%%%%%%%%%%%%%%%%%%%%%%%%%%%%%%%%%%%%%%%%%%%%%%%%%%%%%%%%%%%%

\section*{Acknowledgements} VNŞ thanks Diederik Roest for a useful
discussion.  This work is supported by T\"UB\.ITAK (The Scientific and
Technological Research Council of Turkey) project number 116F385.

%\clearpage

%to prevent pagebreaks in between references:
\makeatletter
\interlinepenalty=10000

%comment out the following lines before submitting to arXiv
%\bibliographystyle{JHEP.bst} 
%\bibliography{yenirefler.bib}

%uncomment the following line before submitting to arXiv
%(skip bibtex and directly use bbl file)
\bibliography{nonminimal_version2}

\providecommand{\href}[2]{#2}\begingroup\raggedright\begin{thebibliography}{10}

\bibitem{Guth:1980zm}
A.~H. Guth, \emph{The inflationary universe: A possible solution to the horizon
  and flatness problems},
  \href{https://doi.org/10.1103/PhysRevD.23.347}{\emph{Phys. Rev.} {\bfseries
  D23} (1981) 347}.

\bibitem{Linde:1981mu}
A.~D. Linde, \emph{A new inflationary universe scenario: A possible solution of
  the horizon, flatness, homogeneity, isotropy and primordial monopole
  problems}, \href{https://doi.org/10.1016/0370-2693(82)91219-9}{\emph{Phys.
  Lett.} {\bfseries 108B} (1982) 389}.

\bibitem{Albrecht:1982wi}
A.~Albrecht and P.~J. Steinhardt, \emph{Cosmology for grand unified theories
  with radiatively induced symmetry breaking},
  \href{https://doi.org/10.1103/PhysRevLett.48.1220}{\emph{Phys. Rev. Lett.}
  {\bfseries 48} (1982) 1220}.

\bibitem{Linde:1983gd}
A.~D. Linde, \emph{Chaotic inflation},
  \href{https://doi.org/10.1016/0370-2693(83)90837-7}{\emph{Phys. Lett.}
  {\bfseries 129B} (1983) 177}.

\bibitem{Ade:2015xua}
{\scshape Planck} collaboration, P.~A.~R. Ade et~al., \emph{Planck 2015
  results. {XIII. Cosmological parameters}},
  \href{https://doi.org/10.1051/0004-6361/201525830}{\emph{Astron. Astrophys.}
  {\bfseries 594} (2016) A13}
  [\href{https://arxiv.org/abs/1502.01589}{{\ttfamily 1502.01589}}].

\bibitem{Ade:2015lrj}
{\scshape Planck} collaboration, P.~A.~R. Ade et~al., \emph{Planck 2015
  results. {XX. Constraints on inflation}},
  \href{https://doi.org/10.1051/0004-6361/201525898}{\emph{Astron. Astrophys.}
  {\bfseries 594} (2016) A20}
  [\href{https://arxiv.org/abs/1502.02114}{{\ttfamily 1502.02114}}].

\bibitem{Martin:2013tda}
J.~Martin, C.~Ringeval and V.~Vennin, \emph{Encyclopædia inflationaris},
  \href{https://doi.org/10.1016/j.dark.2014.01.003}{\emph{Phys. Dark Univ.}
  {\bfseries 5-6} (2014) 75} [\href{https://arxiv.org/abs/1303.3787}{{\ttfamily
  1303.3787}}].

\bibitem{Callan:1970ze}
C.~G. Callan, Jr., S.~R. Coleman and R.~Jackiw, \emph{A new improved energy -
  momentum tensor},
  \href{https://doi.org/10.1016/0003-4916(70)90394-5}{\emph{Annals Phys.}
  {\bfseries 59} (1970) 42}.

\bibitem{Freedman:1974ze}
D.~Z. Freedman and E.~J. Weinberg, \emph{The energy-momentum tensor in scalar
  and gauge field theories},
  \href{https://doi.org/10.1016/0003-4916(74)90040-2}{\emph{Annals Phys.}
  {\bfseries 87} (1974) 354}.

\bibitem{Buchbinder:1992rb}
I.~L. Buchbinder, S.~D. Odintsov and I.~L. Shapiro, \emph{Effective action in
  quantum gravity}. IOP Publishing Ltd, 1992.

\bibitem{Abbott:1981rg}
L.~F. Abbott, \emph{Gravitational effects on the {$SU(5)$} breaking phase
  transition for a {Coleman-Weinberg} potential},
  \href{https://doi.org/10.1016/0550-3213(81)90374-6}{\emph{Nucl. Phys.}
  {\bfseries B185} (1981) 233}.

\bibitem{Spokoiny:1984bd}
B.~L. Spokoiny, \emph{Inflation and generation of perturbations in broken
  symmetric theory of gravity},
  \href{https://doi.org/10.1016/0370-2693(84)90587-2}{\emph{Phys. Lett.}
  {\bfseries 147B} (1984) 39}.

\bibitem{Lucchin:1985ip}
F.~Lucchin, S.~Matarrese and M.~D. Pollock, \emph{Inflation with a nonminimally
  coupled scalar field},
  \href{https://doi.org/10.1016/0370-2693(86)90592-7}{\emph{Phys. Lett.}
  {\bfseries 167B} (1986) 163}.

\bibitem{Futamase:1987ua}
T.~Futamase and K.-i. Maeda, \emph{Chaotic inflationary scenario in models
  having nonminimal coupling with curvature},
  \href{https://doi.org/10.1103/PhysRevD.39.399}{\emph{Phys. Rev.} {\bfseries
  D39} (1989) 399}.

\bibitem{Fakir:1990eg}
R.~Fakir and W.~G. Unruh, \emph{Improvement on cosmological chaotic inflation
  through nonminimal coupling},
  \href{https://doi.org/10.1103/PhysRevD.41.1783}{\emph{Phys. Rev.} {\bfseries
  D41} (1990) 1783}.

\bibitem{Salopek:1988qh}
D.~S. Salopek, J.~R. Bond and J.~M. Bardeen, \emph{Designing density
  fluctuation spectra in inflation},
  \href{https://doi.org/10.1103/PhysRevD.40.1753}{\emph{Phys. Rev.} {\bfseries
  D40} (1989) 1753}.

\bibitem{Amendola:1990nn}
L.~Amendola, M.~Litterio and F.~Occhionero, \emph{The phase space view of
  inflation. 1: {The} nonminimally coupled scalar field},
  \href{https://doi.org/10.1142/S0217751X90001653}{\emph{Int. J. Mod. Phys.}
  {\bfseries A5} (1990) 3861}.

\bibitem{Faraoni:1996rf}
V.~Faraoni, \emph{Nonminimal coupling of the scalar field and inflation},
  \href{https://doi.org/10.1103/PhysRevD.53.6813}{\emph{Phys. Rev.} {\bfseries
  D53} (1996) 6813} [\href{https://arxiv.org/abs/astro-ph/9602111}{{\ttfamily
  astro-ph/9602111}}].

\bibitem{Faraoni:2004pi}
V.~Faraoni, \emph{Cosmology in scalar tensor gravity}. Kluwer Academic
  Publishers, 2004.

\bibitem{Starobinsky:1980te}
A.~A. Starobinsky, \emph{A new type of isotropic cosmological models without
  singularity}, \href{https://doi.org/10.1016/0370-2693(80)90670-X}{\emph{Phys.
  Lett.} {\bfseries 91B} (1980) 99}.

\bibitem{Bauer:2008zj}
F.~Bauer and D.~A. Demir, \emph{Inflation with non-minimal coupling: Metric
  versus {Palatini} formulations},
  \href{https://doi.org/10.1016/j.physletb.2008.06.014}{\emph{Phys. Lett.}
  {\bfseries B665} (2008) 222}
  [\href{https://arxiv.org/abs/0803.2664}{{\ttfamily 0803.2664}}].

\bibitem{Jarv:2017azx}
L.~Järv, A.~Racioppi and T.~Tenkanen, \emph{{Palatini side of inflationary
  attractors}}, \href{https://doi.org/10.1103/PhysRevD.97.083513}{\emph{Phys.
  Rev.} {\bfseries D97} (2018) 083513}
  [\href{https://arxiv.org/abs/1712.08471}{{\ttfamily 1712.08471}}].

\bibitem{Bezrukov:2007ep}
F.~L. Bezrukov and M.~Shaposhnikov, \emph{The standard model {Higgs} boson as
  the inflaton},
  \href{https://doi.org/10.1016/j.physletb.2007.11.072}{\emph{Phys. Lett.}
  {\bfseries B659} (2008) 703}
  [\href{https://arxiv.org/abs/0710.3755}{{\ttfamily 0710.3755}}].

\bibitem{Atkins:2012yn}
M.~Atkins and X.~Calmet, \emph{Bounds on the nonminimal coupling of the {Higgs}
  boson to gravity},
  \href{https://doi.org/10.1103/PhysRevLett.110.051301}{\emph{Phys. Rev. Lett.}
  {\bfseries 110} (2013) 051301}
  [\href{https://arxiv.org/abs/1211.0281}{{\ttfamily 1211.0281}}].

\bibitem{Linde:2011nh}
A.~Linde, M.~Noorbala and A.~Westphal, \emph{Observational consequences of
  chaotic inflation with nonminimal coupling to gravity},
  \href{https://doi.org/10.1088/1475-7516/2011/03/013}{\emph{JCAP} {\bfseries
  1103} (2011) 013} [\href{https://arxiv.org/abs/1101.2652}{{\ttfamily
  1101.2652}}].

\bibitem{Whitt:1984pd}
B.~Whitt, \emph{Fourth order gravity as general relativity plus matter},
  \href{https://doi.org/10.1016/0370-2693(84)90332-0}{\emph{Phys. Lett.}
  {\bfseries 145B} (1984) 176}.

\bibitem{Barbon:2009ya}
J.~L.~F. Barbon and J.~R. Espinosa, \emph{On the naturalness of {Higgs}
  inflation}, \href{https://doi.org/10.1103/PhysRevD.79.081302}{\emph{Phys.
  Rev.} {\bfseries D79} (2009) 081302}
  [\href{https://arxiv.org/abs/0903.0355}{{\ttfamily 0903.0355}}].

\bibitem{Kallosh:2013hoa}
R.~Kallosh and A.~Linde, \emph{Universality class in conformal inflation},
  \href{https://doi.org/10.1088/1475-7516/2013/07/002}{\emph{JCAP} {\bfseries
  1307} (2013) 002} [\href{https://arxiv.org/abs/1306.5220}{{\ttfamily
  1306.5220}}].

\bibitem{Kallosh:2013maa}
R.~Kallosh and A.~Linde, \emph{Non-minimal inflationary attractors},
  \href{https://doi.org/10.1088/1475-7516/2013/10/033}{\emph{JCAP} {\bfseries
  1310} (2013) 033} [\href{https://arxiv.org/abs/1307.7938}{{\ttfamily
  1307.7938}}].

\bibitem{Kallosh:2013tua}
R.~Kallosh, A.~Linde and D.~Roest, \emph{Universal attractor for inflation at
  strong coupling},
  \href{https://doi.org/10.1103/PhysRevLett.112.011303}{\emph{Phys. Rev. Lett.}
  {\bfseries 112} (2014) 011303}
  [\href{https://arxiv.org/abs/1310.3950}{{\ttfamily 1310.3950}}].

\bibitem{Kehagias:2013mya}
A.~Kehagias, A.~Moradinezhad~Dizgah and A.~Riotto, \emph{Remarks on the
  {Starobinsky} model of inflation and its descendants},
  \href{https://doi.org/10.1103/PhysRevD.89.043527}{\emph{Phys. Rev.}
  {\bfseries D89} (2014) 043527}
  [\href{https://arxiv.org/abs/1312.1155}{{\ttfamily 1312.1155}}].

\bibitem{Giudice:2014toa}
G.~F. Giudice and H.~M. Lee, \emph{Starobinsky-like inflation from induced
  gravity}, \href{https://doi.org/10.1016/j.physletb.2014.04.020}{\emph{Phys.
  Lett.} {\bfseries B733} (2014) 58}
  [\href{https://arxiv.org/abs/1402.2129}{{\ttfamily 1402.2129}}].

\bibitem{Galante:2014ifa}
M.~Galante, R.~Kallosh, A.~Linde and D.~Roest, \emph{Unity of cosmological
  inflation attractors},
  \href{https://doi.org/10.1103/PhysRevLett.114.141302}{\emph{Phys. Rev. Lett.}
  {\bfseries 114} (2015) 141302}
  [\href{https://arxiv.org/abs/1412.3797}{{\ttfamily 1412.3797}}].

\bibitem{Mukhanov:1981xt}
V.~F. Mukhanov and G.~V. Chibisov, \emph{Quantum fluctuations and a nonsingular
  universe}, {\emph{JETP Lett.} {\bfseries 33} (1981) 532}.

\bibitem{Okada:2010jf}
N.~Okada, M.~U. Rehman and Q.~Shafi, \emph{Tensor to scalar ratio in
  non-minimal $\phi^4$ inflation},
  \href{https://doi.org/10.1103/PhysRevD.82.043502}{\emph{Phys. Rev.}
  {\bfseries D82} (2010) 043502}
  [\href{https://arxiv.org/abs/1005.5161}{{\ttfamily 1005.5161}}].

\bibitem{Bezrukov:2013fca}
F.~Bezrukov and D.~Gorbunov, \emph{Light inflaton after {LHC8} and {WMAP9}
  results}, \href{https://doi.org/10.1007/JHEP07(2013)140}{\emph{JHEP}
  {\bfseries 07} (2013) 140} [\href{https://arxiv.org/abs/1303.4395}{{\ttfamily
  1303.4395}}].

\bibitem{Fujii:2003pa}
Y.~Fujii and K.~Maeda, \emph{The scalar-tensor theory of gravitation}.
  Cambridge University Press, 2007.

\bibitem{Zee:1978wi}
A.~Zee, \emph{A broken symmetric theory of gravity},
  \href{https://doi.org/10.1103/PhysRevLett.42.417}{\emph{Phys. Rev. Lett.}
  {\bfseries 42} (1979) 417}.

\bibitem{Lyth:2009zz}
D.~H. Lyth and A.~R. Liddle, \emph{The primordial density perturbation:
  {Cosmology,} inflation and the origin of structure}. Cambridge University
  Press, 2009.

\bibitem{Liddle:2003as}
A.~R. Liddle and S.~M. Leach, \emph{How long before the end of inflation were
  observable perturbations produced?},
  \href{https://doi.org/10.1103/PhysRevD.68.103503}{\emph{Phys. Rev.}
  {\bfseries D68} (2003) 103503}
  [\href{https://arxiv.org/abs/astro-ph/0305263}{{\ttfamily
  astro-ph/0305263}}].

\bibitem{Lerner:2009xg}
R.~N. Lerner and J.~McDonald, \emph{Gauge singlet scalar as inflaton and
  thermal relic dark matter},
  \href{https://doi.org/10.1103/PhysRevD.80.123507}{\emph{Phys. Rev.}
  {\bfseries D80} (2009) 123507}
  [\href{https://arxiv.org/abs/0909.0520}{{\ttfamily 0909.0520}}].

\bibitem{Lerner:2009na}
R.~N. Lerner and J.~McDonald, \emph{Higgs inflation and naturalness},
  \href{https://doi.org/10.1088/1475-7516/2010/04/015}{\emph{JCAP} {\bfseries
  1004} (2010) 015} [\href{https://arxiv.org/abs/0912.5463}{{\ttfamily
  0912.5463}}].

\bibitem{Burns:2016ric}
D.~Burns, S.~Karamitsos and A.~Pilaftsis, \emph{Frame-covariant formulation of
  inflation in scalar-curvature theories},
  \href{https://doi.org/10.1016/j.nuclphysb.2016.04.036}{\emph{Nucl. Phys.}
  {\bfseries B907} (2016) 785}
  [\href{https://arxiv.org/abs/1603.03730}{{\ttfamily 1603.03730}}].

\bibitem{Karam:2017zno}
A.~Karam, T.~Pappas and K.~Tamvakis, \emph{Frame-dependence of higher-order
  inflationary observables in scalar-tensor theories},
  \href{https://doi.org/10.1103/PhysRevD.96.064036}{\emph{Phys. Rev.}
  {\bfseries D96} (2017) 064036}
  [\href{https://arxiv.org/abs/1707.00984}{{\ttfamily 1707.00984}}].

\bibitem{Postma:2014vaa}
M.~Postma and M.~Volponi, \emph{{Equivalence of the Einstein and Jordan
  frames}}, \href{https://doi.org/10.1103/PhysRevD.90.103516}{\emph{Phys. Rev.}
  {\bfseries D90} (2014) 103516}
  [\href{https://arxiv.org/abs/1407.6874}{{\ttfamily 1407.6874}}].

\bibitem{Ferrara:2013rsa}
S.~Ferrara, R.~Kallosh, A.~Linde and M.~Porrati, \emph{Minimal supergravity
  models of inflation},
  \href{https://doi.org/10.1103/PhysRevD.88.085038}{\emph{Phys. Rev.}
  {\bfseries D88} (2013) 085038}
  [\href{https://arxiv.org/abs/1307.7696}{{\ttfamily 1307.7696}}].

\bibitem{Kallosh:2013yoa}
R.~Kallosh, A.~Linde and D.~Roest, \emph{Superconformal inflationary
  $\alpha$-attractors},
  \href{https://doi.org/10.1007/JHEP11(2013)198}{\emph{JHEP} {\bfseries 11}
  (2013) 198} [\href{https://arxiv.org/abs/1311.0472}{{\ttfamily 1311.0472}}].

\bibitem{Goldstone:1961eq}
J.~Goldstone, \emph{Field theories with superconductor solutions},
  \href{https://doi.org/10.1007/BF02812722}{\emph{Nuovo Cim.} {\bfseries 19}
  (1961) 154}.

\bibitem{Vilenkin:1994pv}
A.~Vilenkin, \emph{Topological inflation},
  \href{https://doi.org/10.1103/PhysRevLett.72.3137}{\emph{Phys. Rev. Lett.}
  {\bfseries 72} (1994) 3137}
  [\href{https://arxiv.org/abs/hep-th/9402085}{{\ttfamily hep-th/9402085}}].

\bibitem{Linde:1994wt}
A.~D. Linde and D.~A. Linde, \emph{Topological defects as seeds for eternal
  inflation}, \href{https://doi.org/10.1103/PhysRevD.50.2456}{\emph{Phys. Rev.}
  {\bfseries D50} (1994) 2456}
  [\href{https://arxiv.org/abs/hep-th/9402115}{{\ttfamily hep-th/9402115}}].

\bibitem{Destri:2007pv}
C.~Destri, H.~J. de~Vega and N.~G. Sanchez, \emph{{MCMC analysis of WMAP3 and
  SDSS} data points to broken symmetry inflaton potentials and provides a lower
  bound on the tensor to scalar ratio},
  \href{https://doi.org/10.1103/PhysRevD.77.043509}{\emph{Phys. Rev.}
  {\bfseries D77} (2008) 043509}
  [\href{https://arxiv.org/abs/astro-ph/0703417}{{\ttfamily
  astro-ph/0703417}}].

\bibitem{Kallosh:2007wm}
R.~Kallosh and A.~D. Linde, \emph{Testing string theory with {CMB}},
  \href{https://doi.org/10.1088/1475-7516/2007/04/017}{\emph{JCAP} {\bfseries
  0704} (2007) 017} [\href{https://arxiv.org/abs/0704.0647}{{\ttfamily
  0704.0647}}].

\bibitem{Smith:2008pf}
T.~L. Smith, M.~Kamionkowski and A.~Cooray, \emph{The inflationary
  gravitational-wave background and measurements of the scalar spectral index},
  \href{https://doi.org/10.1103/PhysRevD.78.083525}{\emph{Phys. Rev.}
  {\bfseries D78} (2008) 083525}
  [\href{https://arxiv.org/abs/0802.1530}{{\ttfamily 0802.1530}}].

\bibitem{Rehman:2008qs}
M.~U. Rehman, Q.~Shafi and J.~R. Wickman, \emph{{GUT} inflation and proton
  decay after {WMAP5}},
  \href{https://doi.org/10.1103/PhysRevD.78.123516}{\emph{Phys. Rev.}
  {\bfseries D78} (2008) 123516}
  [\href{https://arxiv.org/abs/0810.3625}{{\ttfamily 0810.3625}}].

\bibitem{Okada:2014lxa}
N.~Okada, V.~N. Şenoğuz and Q.~Shafi, \emph{The observational status of
  simple inflationary models: an update},
  \href{https://doi.org/10.3906/fiz-1505-7}{\emph{Turk. J. Phys.} {\bfseries
  40} (2016) 150} [\href{https://arxiv.org/abs/1403.6403}{{\ttfamily
  1403.6403}}].

\bibitem{Ashoorioon:2014jja}
A.~Ashoorioon and M.~M. Sheikh-Jabbari, \emph{{Gauged M-flation after BICEP2}},
  \href{https://doi.org/10.1016/j.physletb.2014.11.018}{\emph{Phys. Lett.}
  {\bfseries B739} (2014) 391}
  [\href{https://arxiv.org/abs/1405.1685}{{\ttfamily 1405.1685}}].

\bibitem{Accetta:1985du}
F.~S. Accetta, D.~J. Zoller and M.~S. Turner, \emph{Induced gravity inflation},
  \href{https://doi.org/10.1103/PhysRevD.31.3046}{\emph{Phys. Rev.} {\bfseries
  D31} (1985) 3046}.

\bibitem{Kaiser:1994vs}
D.~I. Kaiser, \emph{Primordial spectral indices from generalized {Einstein}
  theories}, \href{https://doi.org/10.1103/PhysRevD.52.4295}{\emph{Phys. Rev.}
  {\bfseries D52} (1995) 4295}
  [\href{https://arxiv.org/abs/astro-ph/9408044}{{\ttfamily
  astro-ph/9408044}}].

\bibitem{Cerioni:2009kn}
A.~Cerioni, F.~Finelli, A.~Tronconi and G.~Venturi, \emph{Inflation and
  reheating in induced gravity},
  \href{https://doi.org/10.1016/j.physletb.2009.10.066}{\emph{Phys. Lett.}
  {\bfseries B681} (2009) 383}
  [\href{https://arxiv.org/abs/0906.1902}{{\ttfamily 0906.1902}}].

\bibitem{Tronconi:2017wps}
A.~Tronconi, \emph{Asymptotically safe non-minimal inflation},
  \href{https://doi.org/10.1088/1475-7516/2017/07/015}{\emph{JCAP} {\bfseries
  1707} (2017) 015} [\href{https://arxiv.org/abs/1704.05312}{{\ttfamily
  1704.05312}}].

\bibitem{Izawa:1996dv}
K.~I. Izawa and T.~Yanagida, \emph{Natural new inflation in broken
  supergravity},
  \href{https://doi.org/10.1016/S0370-2693(96)01638-3}{\emph{Phys. Lett.}
  {\bfseries B393} (1997) 331}
  [\href{https://arxiv.org/abs/hep-ph/9608359}{{\ttfamily hep-ph/9608359}}].

\bibitem{Kawasaki:2003zv}
M.~Kawasaki, M.~Yamaguchi and J.~Yokoyama, \emph{Inflation with a running
  spectral index in supergravity},
  \href{https://doi.org/10.1103/PhysRevD.68.023508}{\emph{Phys. Rev.}
  {\bfseries D68} (2003) 023508}
  [\href{https://arxiv.org/abs/hep-ph/0304161}{{\ttfamily hep-ph/0304161}}].

\bibitem{Yamaguchi:2004tn}
M.~Yamaguchi and J.~Yokoyama, \emph{Smooth hybrid inflation in supergravity
  with a running spectral index and early star formation},
  \href{https://doi.org/10.1103/PhysRevD.70.023513}{\emph{Phys. Rev.}
  {\bfseries D70} (2004) 023513}
  [\href{https://arxiv.org/abs/hep-ph/0402282}{{\ttfamily hep-ph/0402282}}].

\bibitem{Senoguz:2004ky}
V.~N. Şenoğuz and Q.~Shafi, \emph{New inflation, preinflation, and
  leptogenesis},
  \href{https://doi.org/10.1016/j.physletb.2004.05.077}{\emph{Phys. Lett.}
  {\bfseries B596} (2004) 8}
  [\href{https://arxiv.org/abs/hep-ph/0403294}{{\ttfamily hep-ph/0403294}}].

\bibitem{Boyanovsky:2007ry}
D.~Boyanovsky, H.~J. de~Vega, C.~M. Ho and N.~G. Sanchez, \emph{New inflation
  vs. chaotic inflation, higher degree potentials and the reconstruction
  program in light of {WMAP3}},
  \href{https://doi.org/10.1103/PhysRevD.75.123504}{\emph{Phys. Rev.}
  {\bfseries D75} (2007) 123504}
  [\href{https://arxiv.org/abs/astro-ph/0702627}{{\ttfamily
  astro-ph/0702627}}].

\bibitem{Destri:2009wn}
C.~Destri, H.~J. de~Vega and N.~G. Sanchez, \emph{{Higher order terms in the
  inflaton potential and the lower bound on the tensor to scalar ratio $r$}},
  \href{https://doi.org/10.1016/j.aop.2010.11.019}{\emph{Annals Phys.}
  {\bfseries 326} (2011) 578}
  [\href{https://arxiv.org/abs/0906.4102}{{\ttfamily 0906.4102}}].

\bibitem{Boubekeur:2005zm}
L.~Boubekeur and D.~H. Lyth, \emph{Hilltop inflation},
  \href{https://doi.org/10.1088/1475-7516/2005/07/010}{\emph{JCAP} {\bfseries
  0507} (2005) 010} [\href{https://arxiv.org/abs/hep-ph/0502047}{{\ttfamily
  hep-ph/0502047}}].

\bibitem{Coleman:1973jx}
S.~R. Coleman and E.~J. Weinberg, \emph{Radiative corrections as the origin of
  spontaneous symmetry breaking},
  \href{https://doi.org/10.1103/PhysRevD.7.1888}{\emph{Phys. Rev.} {\bfseries
  D7} (1973) 1888}.

\bibitem{Shafi:1983bd}
Q.~Shafi and A.~Vilenkin, \emph{Inflation with {$SU(5)$}},
  \href{https://doi.org/10.1103/PhysRevLett.52.691}{\emph{Phys. Rev. Lett.}
  {\bfseries 52} (1984) 691}.

\bibitem{Albrecht:1984qt}
A.~Albrecht and R.~H. Brandenberger, \emph{On the realization of new
  inflation}, \href{https://doi.org/10.1103/PhysRevD.31.1225}{\emph{Phys. Rev.}
  {\bfseries D31} (1985) 1225}.

\bibitem{Linde:2005ht}
A.~D. Linde, \emph{Particle physics and inflationary cosmology},
  {\emph{Contemp. Concepts Phys.} {\bfseries 5} (1990) 1}
  [\href{https://arxiv.org/abs/hep-th/0503203}{{\ttfamily hep-th/0503203}}].

\bibitem{Shafi:2006cs}
Q.~Shafi and V.~N. Şenoğuz, \emph{{Coleman-Weinberg} potential in good
  agreement with {WMAP}},
  \href{https://doi.org/10.1103/PhysRevD.73.127301}{\emph{Phys. Rev.}
  {\bfseries D73} (2006) 127301}
  [\href{https://arxiv.org/abs/astro-ph/0603830}{{\ttfamily
  astro-ph/0603830}}].

\bibitem{Barenboim:2013wra}
G.~Barenboim, E.~J. Chun and H.~M. Lee, \emph{{Coleman-Weinberg} inflation in
  light of {Planck}},
  \href{https://doi.org/10.1016/j.physletb.2014.01.039}{\emph{Phys. Lett.}
  {\bfseries B730} (2014) 81}
  [\href{https://arxiv.org/abs/1309.1695}{{\ttfamily 1309.1695}}].

\bibitem{Kannike:2014mia}
K.~Kannike, A.~Racioppi and M.~Raidal, \emph{{Embedding inflation into the
  Standard Model - more evidence for classical scale invariance}},
  \href{https://doi.org/10.1007/JHEP06(2014)154}{\emph{JHEP} {\bfseries 06}
  (2014) 154} [\href{https://arxiv.org/abs/1405.3987}{{\ttfamily 1405.3987}}].

\bibitem{Senoguz:2015lba}
V.~N. Şenoğuz and Q.~Shafi, \emph{Primordial monopoles, proton decay, gravity
  waves and {GUT} inflation},
  \href{https://doi.org/10.1016/j.physletb.2015.11.037}{\emph{Phys. Lett.}
  {\bfseries B752} (2016) 169}
  [\href{https://arxiv.org/abs/1510.04442}{{\ttfamily 1510.04442}}].

\bibitem{Iso:2014gka}
S.~Iso, K.~Kohri and K.~Shimada, \emph{Small field {Coleman-Weinberg} inflation
  driven by a fermion condensate},
  \href{https://doi.org/10.1103/PhysRevD.91.044006}{\emph{Phys. Rev.}
  {\bfseries D91} (2015) 044006}
  [\href{https://arxiv.org/abs/1408.2339}{{\ttfamily 1408.2339}}].

\bibitem{Kaneta:2017lnj}
K.~Kaneta, O.~Seto and R.~Takahashi, \emph{{Very low scale Coleman-Weinberg
  inflation with nonminimal coupling}},
  \href{https://doi.org/10.1103/PhysRevD.97.063004}{\emph{Phys. Rev.}
  {\bfseries D97} (2018) 063004}
  [\href{https://arxiv.org/abs/1708.06455}{{\ttfamily 1708.06455}}].

\bibitem{Panotopoulos:2014hwa}
G.~Panotopoulos, \emph{{Nonminimal GUT inflation after Planck results}},
  \href{https://doi.org/10.1103/PhysRevD.89.047301}{\emph{Phys. Rev.}
  {\bfseries D89} (2014) 047301}
  [\href{https://arxiv.org/abs/1403.0931}{{\ttfamily 1403.0931}}].

\bibitem{Marzola:2015xbh}
L.~Marzola, A.~Racioppi, M.~Raidal, F.~R. Urban and H.~Veermäe,
  \emph{{Non-minimal CW inflation, electroweak symmetry breaking and the 750
  GeV anomaly}}, \href{https://doi.org/10.1007/JHEP03(2016)190}{\emph{JHEP}
  {\bfseries 03} (2016) 190}
  [\href{https://arxiv.org/abs/1512.09136}{{\ttfamily 1512.09136}}].

\bibitem{Kannike:2015kda}
K.~Kannike, A.~Racioppi and M.~Raidal, \emph{Linear inflation from quartic
  potential}, \href{https://doi.org/10.1007/JHEP01(2016)035}{\emph{JHEP}
  {\bfseries 01} (2016) 035}
  [\href{https://arxiv.org/abs/1509.05423}{{\ttfamily 1509.05423}}].

\bibitem{Baumann:2014nda}
D.~Baumann and L.~McAllister, \emph{Inflation and string theory}. Cambridge
  University Press, 2015, [\href{https://arxiv.org/abs/1404.2601}{{\ttfamily
  1404.2601}}].

\bibitem{Muta:1991mw}
T.~Muta and S.~D. Odintsov, \emph{{Model dependence of the nonminimal scalar
  graviton effective coupling constant in curved space-time}},
  \href{https://doi.org/10.1142/S0217732391004206}{\emph{Mod. Phys. Lett.}
  {\bfseries A6} (1991) 3641}.

\bibitem{Burgess:2009ea}
C.~P. Burgess, H.~M. Lee and M.~Trott, \emph{Power-counting and the validity of
  the classical approximation during inflation},
  \href{https://doi.org/10.1088/1126-6708/2009/09/103}{\emph{JHEP} {\bfseries
  09} (2009) 103} [\href{https://arxiv.org/abs/0902.4465}{{\ttfamily
  0902.4465}}].

\bibitem{Hertzberg:2010dc}
M.~P. Hertzberg, \emph{On inflation with non-minimal coupling},
  \href{https://doi.org/10.1007/JHEP11(2010)023}{\emph{JHEP} {\bfseries 11}
  (2010) 023} [\href{https://arxiv.org/abs/1002.2995}{{\ttfamily 1002.2995}}].

\bibitem{Bezrukov:2010jz}
F.~Bezrukov, A.~Magnin, M.~Shaposhnikov and S.~Sibiryakov, \emph{{Higgs
  inflation: consistency and generalisations}},
  \href{https://doi.org/10.1007/JHEP01(2011)016}{\emph{JHEP} {\bfseries 01}
  (2011) 016} [\href{https://arxiv.org/abs/1008.5157}{{\ttfamily 1008.5157}}].

\end{thebibliography}\endgroup

\makeatother

\end{document}